\documentclass[useAMS,usegraphicx]{mn2e}
\usepackage{graphicx}

\title{The [$\alpha$/Fe] ratios of very metal-poor stars within the
  IGIMF theory}

\author[Recchi et al.]{S.
  Recchi$^{1}$\thanks{simone.recchi@univie.ac.at}, F.  Calura$^{2,
    3}$\thanks{fcalura@oabo.inaf.it}, B.K.  Gibson$^{3,
    4}$\thanks{brad.k.gibson@gmail.com} and P.
  Kroupa$^{5}$\thanks{pavel@astro.uni-bonn.de, 
  Postal address: Argelander Institute for Astronomy (AIfA), Auf dem
  H\"ugel 71, 53121 Bonn, Germany}\\
  $^{1}$ Department of Astrophysics, Vienna University,
  T\"urkenschanzstrasse 17, A-1180, Vienna, Austria \\
  $^{2}$ INAF - Astronomical Observatory of Bologna, Via Ranzani 1,
  40127 Bologna,
  Italy \\
  $^{3}$ Jeremiah Horrocks Institute, University of Central
  Lancashire,
  Preston, PR1 2HE, UK \\
  $^{4}$ Department of Astronomy \& Physics, Saint
  Mary's University, Halifax, Nova Scotia, B3H~3C3, Canada \\
  $^{5}$ Helmholtz-Institut fuer Strahlen und Kernphysik (HISKP), Nussallee 14-16, D-53115 Bonn, Germany}

\date{Received; accepted}

\begin{document}
\maketitle


\begin{abstract}
  The aim of this paper is to quantify the amplitude of the predicted
  plateau in [$\alpha$/Fe] ratios associated with the most metal-poor
  stars of a galaxy.  We assume that the initial mass function in
  galaxies is steeper if the star formation rate (SFR) is low -- as
  per the integrated galactic initial mass function (IGIMF) theory.  A
  variant of the theory, in which the IGIMF depends upon the
  metallicity of the parent galaxy, is also considered. The IGIMF
  theory predicts low [$\alpha$/Fe] plateaus in dwarf galaxies,
  characterised by small SFRs.  The [$\alpha$/Fe] plateau is up to 0.7
  dex lower than the corresponding plateau of the Milky Way.  For a
  universal IMF one should expect instead that the [$\alpha$/Fe]
  plateau is the same for all the galaxies, irrespective of their
  masses or SFRs.  Assuming a strong dependence of the IMF on the
  metallicity of the parent galaxy, dwarf galaxies can show values of
  the [$\alpha$/Fe] plateau similar to those of the Milky Way, and
  almost independent on the SFR.  The [Mg/Fe] ratios of the most
  metal-poor stars in dwarf galaxies satellites of the Milky Way can
  be reproduced either if we consider metallicity-dependent IMFs or if
  the early SFRs of these galaxies were larger than we presently
  think.  Present and future observations of dwarf galaxies can help
  disentangle between these different IGIMF formulations.
\end{abstract}


\maketitle


\section{Introduction}

The relative distribution of stellar masses in a given generation (the
so-called Initial Mass Function or IMF) is a fundamental entity
linking the micro-physics of molecular cloud fragmentation with the
global characteristics of galactic stellar populations.  The IMF
regulates the proportion of high to low mass stars, and hence, the
study of diagnostics such as abundance ratios can constrain its main
features, i.e. its normalisation, upper and lower ends and shape.  The
integrated galactic initial mass function (IGIMF) theory (Kroupa \&
Weidner 2003; Weidner \& Kroupa 2005) is aimed at providing a
self-consistent description of how the IMF varies as a function of
another fundamental characteristic in galaxy evolution models, namely
the star formation (SF) history.  The IGIMF is based on the
fundamental assumptions that ($i$) most stars form in small stellar
groups, i.e. in embedded clusters.  Within each cluster, the stars are
distributed according to a universal two-slope power law IMF:
$\alpha_1=1.3$ below 0.5 M$_\odot$ and $\alpha_2=2.35$ above 0.5
M$_\odot$.  The maximum stellar mass in an individual cluster is a
function of the cluster mass; ($ii$) the stellar clusters are
distributed according to a single-slope power law and ($iii$) the
maximum possible mass of a star cluster increases with the galactic SF
rate.  The main implication is that low mass galaxies, characterised
by low SF values, present a steeper (top-light) IMF than the Milky
Way.

In two previous papers, the IGIMF theory has been applied to study the
abundance ratios in different classes of galaxies.  Recchi et al.
(2009, hereafter R09) investigated the integrated abundances in local
early type galaxies, showing with a chemical evolution model that the
IGIMF theory, coupled with their downsizing character (i.e.  the fact
that the SF timescale inversely correlates with the mass of the
system) well accounts for the observed correlation between integrated
[$\alpha$/Fe] ratios and mass, or velocity dispersion $\sigma$, of the
galaxy.  In particular, these models predict a change in the
[$\alpha$/Fe] vs. $\sigma$ relation not achievable with a constant
IMF.  Calura et al. (2010, hereafter C10) applied the IGIMF to a model
for the solar neighbourhood, showing that a large set of observables
can be accounted for and suggesting how the present-day mass function
can be used locally in order to disentangle between different IMF
scenarios.

In all these studies, the determination of the IMF parameters required
the assumption of a SF history, which is unknown a priori and which
renders the problem degenerate.  In fact, it is well known from
chemical evolution studies that a higher SF efficiency mimics the
effects of a IMF skewed towards massive stars (see e.g. Matteucci
1994).

A different diagnostic able to provide crucial information on the IMF
is the [$\alpha$/Fe] plateau observed at low metallicity in the
Galactic halo (Clegg et al. 1981; Barbuy 1988; Edvardsson et al.
1993).  The importance of the low metallicity plateau is that it
depends solely on the IMF (Tsujimoto et al. 1997), in the sense that
the plateau's value is set entirely by the IMF slope (and massive star
yields), irrespective of the uncertain role played by Type Ia
supernovae.

The situation is different within the IGIMF theory, in that the
plateau depends on the SF history.  If it is possible to establish the
existence of a [$\alpha$/Fe] plateau of metal-poor stars in different
galaxies, its value will depend on the IGIMF slope and on the upper
mass limit, hence it will be a function of the SF history of that
system.  On the other hand, with a constant, SF-independent IMF, all
the galaxies will have the same [$\alpha$/Fe] plateau, irrespective of
the SF history.

To make the previous reasoning more quantitative, let us assume that
all the stars with masses larger than 8 M$_\odot$ explode as Type II
SNe.  It is possible to establish an average [$\alpha$/Fe]$_h$ of the
interstellar medium in the halo as a consequence of SNII pollution
(hence of the halo stars) through the formula:
\begin{equation}
\label{eq:eqgibson}
\left(\frac{\alpha}{Fe}\right)_h=\left(\frac{\alpha}{Fe}\right)_\odot 
\times 10^{[\alpha/Fe]_h}=\frac{\int_{8}^{m_{up}}m_\alpha^{ej}
\varphi(m)dm}{\int_{8}^{m_{up}}m_{Fe}^{ej}
\varphi(m)dm},
\end{equation}
\noindent
where $\varphi(m)$ is the adopted IMF, $m_{up}$ is its upper limit and
$m_\alpha^{ej}$, $m_{Fe}^{ej}$ are the theoretical yields from Type II
SNe (see Tsujimoto et al. 1997; Gibson 1998).  Once a compilation of
SNII yields is established (for instance the widely used Woosley \&
Weaver 1995 compilation), the [$\alpha$/Fe] plateau is determined
solely by the IMF, in particular by its slope at the high-mass range
and upper limit.  Tsujimoto et al. (1997) showed in this way that the
[O/Fe] in Galactic metal-poor halo stars is consistent with an IMF
with slope 2.3--2.6 above 1 M$_\odot$ (i.e.  similar or slightly
larger than the Salpeter index) and an upper mass limit $m_{up}=50\pm
10$ M$_\odot$.

Although Gibson (1998) pointed out all the limits and weaknesses of
this approach, nevertheless a point should be made.  If an
[$\alpha$/Fe] plateau of metal-poor stars in different galaxies is
established, its value can only reflect the IMF slope(s) and upper
mass limit.  The yields of massive stars, although still very
uncertain, cannot change from galaxy to galaxy\footnote{Actually, one
  might be able to envision some scenario where yields could differ
  from galaxy to galaxy, for instance the fraction of high rotators
  might differ as a function of environment.  However, this hypothesis
  is very speculative at this stage.}.  If the IMF is the same in all
galaxies, the most metal-poor stars must have on average the same
[$\alpha$/Fe], irrespective of the galaxy they belong to.  If instead
a variation of the [$\alpha$/Fe] plateau among different galaxies is
established, this is a clear indication that the IMF is not universal.

This provides a potentially interesting test for the IGIMF theory
because the latter predicts that dwarf galaxies, characterised by
average mild-low SF rates (SFRs), should have steeper IMFs (and lower
values of $m_{up}$) than the Milky Way (Weidner \& Kroupa 2005).  At
variance with the previously published works of R09 and C10, this test
for the IGIMF is independent of the weakly constrained role of Type Ia
SNe and low- and intermediate-mass stellar yields on the abundance
patterns.

It is worth pointing out that the IMF variations we are dealing with
here are concentrated in the upper range of stellar masses. The IGIMF
theory predicts little changes in the IMF for stars with low and
intermediate masses, whereas the change of the IMF slope for high-mass
stars and of the upper mass cut off can be extremely significant (see
e.g. Fig. 1 of R09 or Fig. \ref{igimf} in this paper).  The change of
the IMF in this mass range can directly and significantly affect the
[$\alpha$/Fe] plateaus.  Other forms of IMF variations among galaxies
has been inferred/deduced recently.  On the one hand, it seems that
also the low-mass end of the IMF changes according to the galaxy mass
(Conroy \& van Dokkum 2012; Ferreras et al.  2013).  On the other
hand, also the metallicity can have a significant role in the star
formation process, regulating the transition between PopIII and PopII
stars, whose IMFs are supposed to be very different (see e.g.  Bromm
et al.  2001; Schneider et al. 2002.  See also Sect.  \ref{sec:disc}
for further discussions on the dependence of the IMF on metallicity).

The paper is organised as follows: In Sec. \ref{sec:igimf} the main
assumptions behind the IGIMF theory will be shortly summarised; in
Sect. \ref{sec:afe_obs} we summarise the present knowledge regarding
[$\alpha$/Fe] plateaus in dwarf galaxies.  Given the wealth of
available data, much of this section will be devoted to the abundance
patterns of individual stars in the most luminous dwarf galaxy
satellites (DGSs) of the Milky Way (MW), namely the ones for which we
know spectroscopic data of single stars with high precision.  In Sect.
\ref{sec:sfh_obs} we will also summarise our present knowledge
regarding the SFRs and SF histories of these galaxies, given the key
role that SF histories play in the IGIMF theory.  In Sect.
\ref{sec:calc} the main assumptions and the calculations required to
predict the [$\alpha$/Fe] plateaus of our model galaxies will be
outlined.  In Sect. \ref{sec:res} the results of our investigation
will be presented.  We will concentrate first (Sect.  \ref{sec:resmw})
on the study of the Galactic Milky Way plateau, in order to identify
from the literature the most accurate value of the [$\alpha$/Fe]
plateau and to calculate the best set of nucleosynthetic yields able
to reproduce it with our simplified approach.  In Sect.
\ref{sec:resgal} we will present results concerning the variations in
the [$\alpha$/Fe] plateau as a function of the SFR that we predict
with our model.  In Sect.  \ref{sec:zdepigimf}, results obtained by
using new, metallicity-dependent IGIMF formulations (Marks et al.
2012, hereafter M12) are presented.  In Sect.  \ref{sec:compwithobs},
we compare our results with available observations.  Finally, in Sect.
\ref{sec:conc}, some conclusions are drawn.

\section{The integrated galactic initial mass function}
\label{sec:igimf} 
In this Section, we briefly summarise the main assumptions behind the
IGIMF theory, already shortly anticipated in the Introduction.  A more
complete description is provided in Weidner \& Kroupa (2005) and R09.

The main assumption of the IGIMF theory is that all the stars in a
galaxy form in star clusters.  These star clusters need not be
gravitationally bound entities, but should be seen as space-time
correlated stellar populations, deeply embedded in gas.  These
embedded clusters have, prior to the removal of their remaining gas,
radii of about 0.15 pc, in agreement with molecular cloud cores and
the filamentary structures of molecular clouds (Andr{\'e} et al.
2010; Hennemann et al. 2012; Schneider et al.  2012; Marks \& Kroupa
2012)\footnote{Some authors (e.g. Bressert et al. 2010) claim that a
  significant fraction of massive stars might have formed far from
  dense environments.  It is however worth noticing that a log-normal
  surface density distribution (taken by Bressert et al. as an
  evidence of significant star formation far from dense environments)
  can be achieved also by means of stars forming in bound clusters
  (see e.g.  Gieles et al. 2012).}.  Within each embedded cluster, the
stellar IMF has the canonical form $\xi(m) = k m^{-\alpha}$, with
$\alpha_1 = 1.3$ for 0.08 M$_\odot \le$ $m$ $<$ 0.5 M$_\odot$ and
$\alpha_2 = 2.35$ (i.e.  the Salpeter slope) for 0.5 M$_\odot \le$ $m$
$< m_{\rm max}$.  The upper mass $m_{\rm max}$ is a function of the
mass of the embedded cluster $M_{\rm ecl}$; this is simply due to the
fact that small clusters do not have enough mass to produce very
massive stars.

On the other hand, the mass distribution function of the correlated
star formation entities (embedded clusters) is assumed to be a
single-slope power law, $\xi_{\rm ecl} \propto M_{\rm ecl}^{-\beta}$,
with a slope $\beta$ close to 2 (Zhang \& Fall 1999; Lada \& Lada
2003).  In this work, as in our previous IGIMF studies, we have
considered values of $\beta$: 1.00, 2.00 and 2.35.  According to the
quoted papers and to the results of R09 and C10 the reference value
for $\beta$ will be 2, but it is of interest to evaluate the
dependence of our results on the choice of $\beta$.  The combination
of the stellar IMF with the mass distribution of embedded clusters
yields the IGIMF, i.e.  the IMF integrated over the whole population
of embedded clusters forming in a galaxy as a function of the star
formation rate $\psi(t)$:

\begin{equation}
\xi_{\rm IGIMF}(m;{\psi} (t)) = 
\int_{M_{\rm ecl, min}}^{M_{\rm ecl, max} ({\psi} (t))} 
\hspace{-0.6cm}\xi (m \leq m_{\rm max}) \xi_{\rm ecl} (M_{\rm ecl}) 
dM_{\rm ecl}, 
\label{eq:igimf}
\end{equation}
where $M_{\rm ecl, min}$ and $M_{\rm ecl, max} (\psi (t))$ are the
minimum and maximum possible masses of the embedded clusters in a
population of clusters, respectively, and $m_{\rm max} = m_{\rm max}
(M_{\rm ecl})$.  For $M_{\rm ecl, min}$ we assume the value of 5
M$_\odot$, i.e. the mass of a Taurus-Auriga aggregate, which is
arguably the smallest star-forming "cluster" known.  On the other
hand, the upper mass of the embedded cluster population depends on the
SFR; this fact renders the whole IGIMF dependent on $\psi$.  The
existence of a correlation between $M_{\rm ecl, max}$ and SFR has been
observationally established (Larsen \& Richtler 2000; Weidner et al.
2004) and originates from the sampling of clusters from the embedded
cluster mass function given the amount of gas mass being turned into
stars per unit time (Weidner et al. 2004).  In accordance with Weidner
et al. (2004), we assume this function to be:
\begin{equation}
\log M_{\rm ecl, max} = A + B \log \psi,
\end{equation}
\noindent
with $A=4.83$ and $B=0.75$.

\begin{figure}
\centering
\vspace{0.001cm}
\includegraphics[width=9cm]{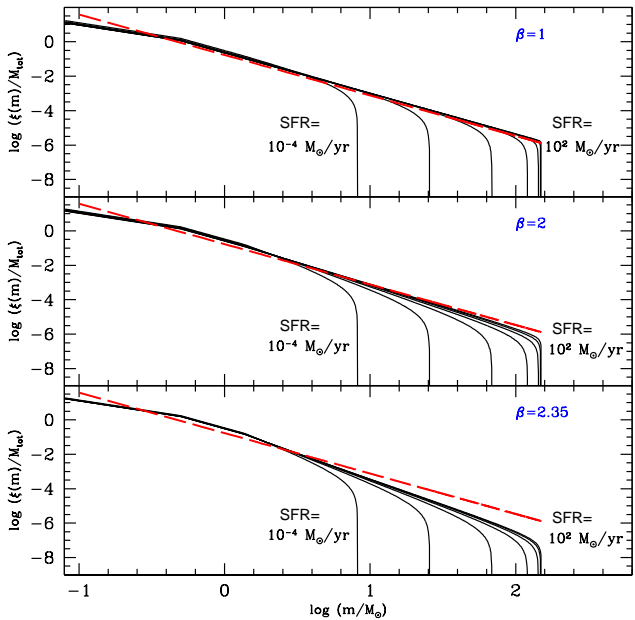}
\caption{IGIMFs (in logarithm) for different cluster mass functions
  distributions.  Upper panel: $\beta=1$; central panel: $\beta=2$;
  lower panel: $\beta=2.35$.  In each panel we have considered 7
  possible values of SFRs, ranging from $10^{-4}$ M$_{\odot}$
  yr$^{-1}$ (lower-most solid lines) to 100 M$_{\odot}$ yr$^{-1}$
  (uppermost solid lines), equally spaced in logarithm. The thick
  dashed line is the standard Salpeter IMF.  The IGIMFs and the
  Salpeter IMF are all normalised through the equation $\int m \xi (m)
  dm=M_{tot}=1M_{\odot}$.  }
\label{igimf}
\end{figure}

In Fig.~\ref{igimf}, we show the IGIMF as a function of the SFR for
the three values of $\beta$ considered in this work, compared to a
single-slope, Salpeter IMF.  The IGIMFs are characterised by a nearly
uniform decline, nearly approximated by a power law, and a sharp
cutoff at mass values close to $m_{\rm max}$.  The steepest
distribution of embedded cluster (in our case the model with
$\beta=2.35$) produces also the steepest IGIMF; this occurs since this
distribution is biased towards embedded clusters of low mass,
therefore the probability of finding high mass stars in this cluster
population is lower.  Moreover, the dependence of the IGIMF on the SFR
is strong for SFR $\le$ 1 M$_\odot$ yr$^{-1}$ whereas it is weak for
SFR $\ge$ 1 M$_\odot$ yr$^{-1}$ (see R09).  This is due to the fact
that for SFRs larger than $\sim$ 1 M$_\odot$ yr$^{-1}$, the maximum
possible value of the embedded cluster mass is very high, therefore it
is always possible to sample massive stars in a galaxy up to a mass
very close to the empirical limit (which is assumed to be $m_{\rm max,
  *}=$ 150 M$_\odot$; see Weidner \& Kroupa 2005).

It is worth noticing that here the IMF within each embedded cluster is
assumed to always be invariant.  That is, neither the possibly of
having a top-heavy IMF in star-burst clusters (which leads to a
top-heavy IGIMF for SFR $>$ 10 M$_\odot$ yr$^{-1}$; see  
Weidner et al. 2011; Kroupa et al.  2013), nor the alleged
dependence of the IMF in clusters on the metallicity (Kroupa et al.
2013; M12) are incorporated here.  The first hypothesis will not be
considered in detail since this paper is mostly concerned with dwarf
galaxies, whose SFRs cannot reach high values.  We will instead study
in Sect.  \ref{sec:zdepigimf} the possibility of having a dependence
of the IMF on the metal content of the parent galaxy.  We remark
already, however, that M12 mostly based their conclusions on the study
of the mass distribution functions in globular clusters.  It is not
clear at this stage if this model can be safely extended to entire
galaxies.

A summary of the main IGIMF parameters and their reference values used
in this study is provided in Table ~\ref{table1}, whereas the main
chemical evolution parameters assumed in the study of the Milky Way
(Sect. \ref{sec:resmw}) are reported in Table ~\ref{table2}.

\begin{table*}
\caption{IGIMF parameters and their reference values used in this study}
\label{table1}      
\centering                          
\begin{tabular}{c c c c c}        
\hline\hline                 
Quantity         &  Adopted Values &    &    Ref.  & Sect. \\ 
\hline                        
Constant & & & & \\
stellar IMF& $\xi(m) = k m^{-\alpha}$  &  $\alpha_{1} = 1.30$; &   
Weidner \& Kroupa (2005)     &  ~\ref{sec:igimf}     \\ 
(IMF within a & &  0.08 M$_\odot \le$ $m$ $<$ 0.5 M$_\odot$  &  & \\
star cluster) & &  $\alpha_{2} = 2.35$; & & \\ 
 & &  0.5 M$_\odot \le$ $m$ $< m_{\rm max}$ & &\\
\hline
Empirical stellar & & & & \\
mass limit &  $m_{\rm max, *}= 150 M_{\odot}$ & 
($m_{\rm max} < m_{\rm max, *}$) & Weidner \& Kroupa 
(2005)   & ~\ref{sec:igimf} \\
\hline
Cluster Mass & & & &\\
Function &   $\xi_{\rm ecl} \propto M_{\rm ecl}^{-\beta}$& 
$\beta =2$  & Zhang \& Fall (1999)   & ~\ref{sec:igimf}   \\
\hline
Minimum & & & &     \\
cluster mass &     $M_{\rm ecl, min}=$ 5 M$_\odot$ & & 
Weidner \& Kroupa (2005)   & ~\ref{sec:igimf} \\
\hline
Maximum & & & &     \\
cluster mass &     $M_{\rm ecl, max}= A + B \log \psi$ & $A=4.83$, $B=0.75$& 
Weidner \& Kroupa (2005)   & ~\ref{sec:igimf} \\
\hline                                   
\end{tabular}
\end{table*}

\section{The satellites of the Milky Way as a benchmark of the IGIMF
  theory}
\label{sec:satellites}

The best benchmark to test possible variations in the [$\alpha$/Fe]
plateau among different galaxies and link it to the variations of the
IMF is the Local Group, with particular emphasis on the MW DGSs.  Most
of these dwarf galaxies (in particular the {\it classical} satellites;
Kroupa et al.  2010) are very well known and studied, both
photometrically and spectroscopically.

\subsection{The [$\alpha$/Fe] plateau in DGSs}
\label{sec:afe_obs}
A wealth of data concerning the chemical composition of individual
stars in DGSs is now available (see for instance Kirby et al. 2011
and references therein).

Useful data concerning [$\alpha$/Fe] ratios in DGSs come from the DART
(Dwarf Abundances and Radial Velocities Team) experiment (Tolstoy et
al. 2004), which is dedicated to the measure of abundances and
velocities for several hundred individual stars in a sample of four
nearby (luminous) dwarf galaxies: Sculptor, Fornax, Sextans and
Carina.  Among these galaxies, perhaps the best example of a
[$\alpha$/Fe] halo plateau is given by Fornax (Letarte et al.  2010).
Looking at the [Mg/Fe] ratios in Fornax, there does not seem to be a
change in [$\alpha$/Fe] for stars more metal-poor than [Fe/H]=$-1$ and
the average value is about 0.1 dex.  Only one star in their sample has
[$\alpha$/Fe] $\simeq$ 0.4; the other [$\alpha$/Fe] ratios are all
close to solar (see also de Boer et al. 2012b).  In addition, globular
clusters in Fornax have an average [Mg/Fe]=0.025 (Larsen et al. 2012).
In Sextans, too, most of the observed stars have [Mg/Fe] ratios close
to the solar value (Aoki et al. 2009).

Also Sagittarius seems to lack (or to have very few) metal-poor stars
with the same [Mg/Fe] of metal-poor halo stars of the MW, although in
this case the establishment of a plateau is much more uncertain (see
Tolstoy et al. 2009).  Moreover, recently McWilliam et al. (2013)
analyzed three very $\alpha$-deficient stars in Sagittarius and
concluded that their abundance ratios can only be compatible with an
IMF very deficient in massive stars.  They thus argued that the
abundance ratios in Sagittarius can be well explained by the IGIMF
theory.  Carina shows large dispersions of [Mg/Fe] ratios, perhaps
indicative of inhomogeneous mixing of supernovae ejecta (Venn et al.
2012).  A more recent, larger sample of Carina stars (Lemasle et al.
2012) seems also to indicate that the [Mg/Fe] of metal-poor stars is
not far from the Milky Way plateau.  However, if we consider only the
stars with [Fe/H] $<-2$, the average [Mg/Fe] ratio of the Lemasle
sample is close to zero (see Sect.  \ref{sec:resgal}).  Moreover,
looking at the [Ca/Fe] ratios of all stars older than 10 Gyr, the
average [$\alpha$/Fe] is less than that observed in the Galaxy.
Sculptor instead seems to have a distribution of [Mg/Fe] ratios in
metal-poor stars similar to the Galactic one (de Boer et al. 2012a).
Data about the [Mg/Fe] ratios in the most metal-poor stars of Draco,
Carina, Sextans and Sculptor are further discussed and compared with
model results in Sect.  \ref{sec:resgal}.

Medium-resolution spectra of about 3000 stars in a larger sample of
DGSs, including ultra-faint ones are also available (Kirby et al.
2010; see also Kirby et al. 2011).  This study challenges some of the
conclusions of the DART experiment in the sense that the average
[Mg/Fe] of metal-poor ([Fe/H]$<$-1.8) stars in DGSs seems to be even
slightly higher than that of the MW.  Moreover, they do not detect any
[$\alpha$/Fe] plateau for [Fe/H]$>$-2.5, with maybe the exception of
Sculptor, and conclude that SNeIa, responsible for the production of
the bulk of iron, substantially contributed to the chemical evolution
even at these low metallicities.  Also the analysis of Frebel et al.
(2010) of the ultra-faint dwarf galaxies UMa II and Coma Berenices,
indicates either the presence of a plateau with relatively high
[$\alpha$/Fe] or a constant decrease of [$\alpha$/Fe] with [Fe/H],
i.e. no plateau (see also Vargas et al. 2013).  The ultra-faint dwarf
galaxy Leo IV has been analysed, too, (Simon et al. 2010) and, again,
no compelling signs of a [Mg/Fe] plateau significantly lower than the
Galactic one have been found.  To finish, Norris et al. (2010) analyse
a star in Bootes I with [Fe/H] = -3.7 and high [$\alpha$/Fe].

As a summary of this Section, only a fraction of the DGSs show signs
of an enrichment of $\alpha$-elements in metal-poor stars
significantly lower than the one observed in the Milky Way, whereas
some other DGSs seem to show either [$\alpha$/Fe] plateaus similar to
the ones found in the MW, or no signs of plateau at all.  Of course,
deeper and more detailed observations might change this picture.  It
is also worth pointing out that non-LTE effects can significantly
change the measurements of the abundances of individual stars, as
shown by Fabrizio et al.  (2012) in the case of Carina.  In spite of
the incredible progresses made in the last years, the study of the
abundances of individual stars in dwarf galaxies is still in its
infancy.

\subsection{The star formation histories of DGSs}
\label{sec:sfh_obs}

In many cases, the whole SF history of DGSs has been recovered by
means of comparison of synthetic color-magnitude diagrams (CMDs) with
observed ones (see Tolstoy et al. 2009 and references therein).
Particularly relevant in this sense is the LCID project (Monelli et
al. 2007), aimed at obtaining CMDs reaching the oldest main-sequence
turnoff stars for a sample of nearby isolated dwarf galaxies, using
the ACS camera on board of the Hubble Space Telescope.  The LCID team
has already obtained detailed SFRs for the dwarf galaxies LGS-3,
Tucana, Cetus, Phoenix (see Monelli et al.  2010a, 2010b; Hidalgo et
al. 2011).

It would be extremely interesting to compare the SFRs in the {\it
  early} phases of the evolution of these galaxies, when the
contribution of SNeIa to the chemical evolution is supposed to be
negligible, and to compare it with the [$\alpha$/Fe] plateau (if any)
in the most metal-poor stars.  These two quantities should correlate.
However, tidal interactions and other environmental effects can be
very strong in DGSs and this could lead to the loss of a very
significant fraction of the baryonic mass initially present in the
galaxy.  In the extreme cases, the mass of a galaxy can be reduced by
two or three orders of magnitude (Kroupa 1997; Klessen \& Kroupa 1998;
Mayer et al. 2006).  Therefore, the evaluation of the SF history of
DGSs made for instance by the LCID team by means of the analysis of
the present-day CMD cannot correctly take into account the large
fraction of long-living stars that the galaxy has lost orbiting around
the MW.  The present-day CMDs of DGSs contain only the (long-living)
stars that survived $\sim$ 12 Gyr of galactic evolution; they do not
contain all long-living stars ever formed in the galaxy.  Of course,
if some of these stars got stripped off the galaxy the CMDs will not
contain them, and the resulting SFR determinations will necessarily be
lower limits.

In some cases, there is direct observational evidence of tidal
disruption of DGSs.  The archetypal example is Sagittarius (Ibata et
al. 1994; Majewski et al. 2003), but also Segue 2 is a bare remnant of
a tidally stripped galaxy (Kirby et al. 2013).  Other DGSs like Draco
(Cioni \& Habing 2005) and Ursa Minor (Palma et al. 2003) clearly show
S-shaped morphologies, indicative of tidal interactions.  It is also
important to point out that modern techniques of reconstruction of the
SF histories of galaxies still do not have adequate resolution for
very old stars and an ancient SF peak, if present, would be smeared
out and the deduced peak SFR could be under-estimated by a factor of
up to 10 (see de Boer et al. 2012b, their Fig. 10).  Therefore, only
the {\it relative} SF histories deduced by these authors, i.e.  the
fraction of stars formed during different periods of the galactic
evolution, should be considered reliable (see also the discussion
about SF histories in DGSs in Sect.  \ref{sec:compwithobs}).  Real
SFRs in the early evolution of these galaxies might have been 2 orders
of magnitude larger than commonly reported.

For later convenience, we summarise here some typical derived early
SFRs in DGSs, bearing in mind the caveats expressed above.  Results of
the LCID project suggest that the SFRs of dwarf galaxies, even the
ones for which a short episode of SF is deduced (for instance Cetus;
Monelli et al. 2010b), are mild, often below 10$^{-4}$ M$_\odot$
yr$^{-1}$, although the SFR of Tucana might have been comparable to
the SFR deduced in present-day blue compact dwarfs.  Early SF
histories with similar SFRs are suggested also for Sagittarius
(Karachentsev et al. 1999) and Fornax (Coleman \& De Jong 2008),
although the early SFR in Fornax might have been as high as 10$^{-3}$
M$_\odot$ yr$^{-1}$ (de Boer et al.  2012b).  On the theoretical side,
the models of Salvadori et al.  (2008) indicate (very short) early
bursts of SF in DGSs with SFRs of at least 0.1--0.15 M$_\odot$
yr$^{-1}$ (but see Lanfranchi et al. 2006; Fenner et al. 2006; Calura
\& Menci 2009; Brook et al. 2012).

\section{The galaxy model}
\label{sec:calc}
We want to establish the correlation between the IMF of a galaxy and
its [$\alpha$/Fe] plateau (or at least the average [$\alpha$/Fe] of
the most metal-poor stars) using an approach similar to the one used
by Tsujimoto et al. (1997) (namely, by means of Eq.
\ref{eq:eqgibson}), but in the framework of the IGIMF theory.

In order to do that, we make the following assumptions:
\begin{itemize}
\item All the stars above a threshold mass $m_{thr}$ explode
  instantaneously as Type II SNe
\item The stars with masses below $m_{thr}$ do not contribute to the
  chemical enrichment.  This assumption is always justified for
  $\alpha$-elements whereas it is not justified, in general, for iron.
  In fact, as already mentioned, a large fraction of iron comes from
  Type Ia SNe, whose progenitors have non-negligible lifetimes.
  However, by definition [$\alpha$/Fe] plateaus are regions in the
  [Fe/H]-[$\alpha$/Fe] plane where the contribution of SNeIa can be
  neglected (a non-negligible contribution from SNeIa would decrease
  the [$\alpha$/Fe] as a function of [Fe/H]).  Hence, provided that we
  can really identify [$\alpha$/Fe] plateaus in galaxies (see Sect.
  \ref{sec:afe_obs}), our assumption is fully reasonable.
\item The chemical products of SNeII are instantaneously and
  homogeneously mixed with the surrounding interstellar medium.
  Although this assumption is not fully justified theoretically, it is
  the only possible assumption in the present framework.  This
  hypothesis can be relaxed only by means of detailed chemo-dynamical
  simulations (in progress).  Results of Spitoni et al.  (2009) show
  that the inclusion of a delayed mixing does not dramatically alter
  the [$\alpha$/Fe] tracks.
\item The initial metallicity of the galaxy is set to be Z=10$^{-4}$
  Z$_\odot$.  This assumption is further discussed in Sect.
  \ref{sec:resmw}.
\end{itemize}

The [$\alpha$/Fe] plateaus in galaxies as defined in Eq.
\ref{eq:eqgibson}, but with the the universal IMF $\varphi(m)$
replaced by the ($\psi$-dependent) IGIMF $\xi_{\rm IGIMF}$ is defined
as:
\begin{equation}
  \label{eq:plateau}
  [\alpha/Fe]_{pl}=\log \left[ \frac{\int_{m_{thr}}^{m_{up}}m_\alpha^{ej}
      \xi_{\rm IGIMF}(m,\psi)dm}{\int_{m_{thr}}^{m_{up}}m_{Fe}^{ej}
      \xi_{\rm IGIMF}(m,\psi)dm}\right] - \log \left( \frac{\alpha}{Fe}
  \right)_{\odot},
\end{equation}
\noindent
where $m_{up}$ in this case is the upper mass in the largest star
cluster of the galaxy (i.e. $m_{up}=m_{\rm max}\left[M_{\rm ecl, max}
  (\psi)\right]$) and [$\alpha$/Fe]$_{pl}$ indicates the [$\alpha$/Fe]
plateau we seek and, as we can see, it depends on the SFR $\psi$.

\begin{table*}
  \caption{Milky Way chemical evolution parameters and reference values 
    used in this study}
\label{table2}      
\centering                          
\begin{tabular}{c c c c }        
\hline\hline                 
Quantity  &  Adopted Values   & Ref.                    & Sect. \\ 
\hline
$<SFR>_{1 Gyr}$ ($M_{\odot}/yr$) & 0.5 &  C10& 
\ref{sec:resmw}   \\
$[$O/Fe$]$ &  0.47 $\pm$ 0.15 &  Cayrel et al. (2004) & 
\ref{sec:resmw}  \\
$[$Si/Fe$]$ &  0.37 $\pm$ 0.15 &  Cayrel et al. (2004) & 
\ref{sec:resmw}  \\
$[$Mg/Fe$]$ &  0.27 $\pm$ 0.13 &  Cayrel et al. (2004) & 
\ref{sec:resmw}  \\
 ${\hat y}_{Fe}$ & 0 & Fig. \ref{fig:fit2} & \ref{sec:resmw}  \\
 $m_{thr}$ & 8 M$_\odot$ & - & \ref{sec:resmw} \\
 $y_{Fe, 150}/y_{Fe, 40}$ & 1 & - & \ref{sec:resmw} \\
\hline                                   
\end{tabular}
\end{table*}

\section{Results}
\label{sec:res}
\subsection{The optimal set of yields to reproduce the Milky Way 
[$\alpha$/Fe] plateau}
\label{sec:resmw}
We wish to apply Eq. \ref{eq:plateau} and calculate
[$\alpha$/Fe]$_{pl}$ in model galaxies characterised by different
values of the SFR $\psi$.  In order to do that, we first need to check
if such an approach allows us to reproduce the MW [$\alpha$/Fe]
plateau.  We thus apply Eq. \ref{eq:plateau} for a value of the SFR
characterising the early phases (for instance the first Gyr) of the MW
evolution.  In particular, we take $<SFR>_{1 Gyr}\simeq$ 0.5 M$_\odot$
yr$^{-1}$.  This value comes from the calculations of C10 of the
chemical evolution of the Solar Neighbourhood.  It is important to
remark that, in the work of C10, the results are calibrated on the
empirical data available for the inner halo of the Milky Way; at
present, there is insufficient high-resolution data available for
field stars of the outer halo where the impact of individual accretion
events on the position/tightness of the plateau might be more
important.  However, Carollo et al. (2010) argue that the inner and
outer halos have mean [Fe/H] = -1.6 and -2.2, respectively.  If this
is the case, our emphasis of stars with [Fe/H] $<$ -2.0 in the solar
neighbourhood could be prone to contamination by the outer halo.  It
is important to point out that our results are quite insensitive to
the choice of $<SFR>_{1 Gyr}$ (see Appendix \ref{sec:app}), hence we
keep $<SFR>_{1 Gyr}=$ 0.5 M$_\odot$ yr$^{-1}$ as our reference value.
This value, together with all other values adopted as standard values
for our MW plateau calculation, are reported in Table \ref{table2}.
We wish to find out for which set of yields Eq. \ref{eq:plateau}
correctly reproduces the abundance ratios observed in the most
metal-poor MW stars.  This set of yields can be safely applied to
other model galaxies, since we do not expect galaxy-to-galaxy yields
variations (see also the Introduction).

The reference dataset we will use for very metal-poor stars in the
Milky Way halo comes from the Cayrel et al. (2004) paper.  This
dataset is sufficiently rich, uniform and possesses little intrinsic
scatter.  It seems to be the best reference to obtain a value for the
observed [$\alpha$/Fe] MW plateau.  The four most commonly (and best)
measured $\alpha$-elements in galaxies are O, Mg, Si and Ca.  Calcium
is perhaps the best measured element among these four because it has
more lines that can be measured than the other $\alpha$-elements,
especially in DGSs.  The dispersion of the [Mg/Fe] ratios observed by
Cayrel et al. (2004) about the lines of the best fit is also smaller
than the dispersion of other $\alpha$-elements.  However, Ca yields
are very uncertain (much more than O, Mg and Si yields) because they
are strongly affected by explosive burning and ``fallback'' (Woosley
\& Weaver 1995, hereafter WW95; Shigeyama \& Tsujimoto 1998).  This
makes the comparison between models and observations very problematic
and uncertain.  In this sense, Mg is a much more robust
$\alpha$-element because its production is dominated by hydrostatic
burning processes and it has been often used as ``reference element''
(Shigeyama \& Tsujimoto 1998; Cayrel et al.  2004).  Moreover, Type Ia
SNe contribute substantially to the production of Ca.  According to
Iwamoto et al.  (1999, see their Table 3), a typical (IMF averaged)
SNII produces 5.8 $\cdot$ 10$^{-3}$ M$_\odot$ of Ca, whereas the
(SNIa) W7 model predicts a yield of 1.2 $\cdot$ 10$^{-2}$ M$_\odot$,
i.e. Type Ia SNe are responsible for the majority of the Ca
production.  On the other hand, the average Mg from SNeII is 8.8
$\cdot$ 10$^{-2}$ M$_\odot$, whereas the Ia production of Mg is only
8.5$\cdot$ 10$^{-3}$ M$_\odot$ (and also Mg from intermediate-mass
stars is negligible).  Finally, the slope of the [Ca/Fe] vs. [Fe/H]
correlation identified by Cayrel et al.  (2004) is not as flat as for
the other $\alpha$-elements (see Table 9 of Cayrel et al. 2004).
Given the large number of Mg abundance measurements in individual
stars of DGSs (see also Sect.  \ref{sec:resgal}), we will use Mg as a
reference element throughout this paper, although we will consider
also O, Si and, occasionally, Ca.  In what follows, we will assume
[Si/Fe]=0.37 $\pm$ 0.15, [O/Fe]=0.47 $\pm$ 0.15 and [Mg/Fe]=0.27 $\pm$
0.13 as typical values of the [$\alpha$/Fe] MW plateau. Cayrel et al.
(2004) report also the measurements of other $\alpha$-elements.  The
inclusion of other, less accurately measured, $\alpha$-elements (S, Ar
or Ti) adds nothing to the conclusions of this paper.  We will assume
solar abundances from Asplund et al.  (2009), namely 12+log(O/H)=8.69,
12+log(Si/H)=7.51, 12+log(Mg/H)=7.60, 12+log(Fe/H)=7.50.

As shown by C10, chemical evolution MW models based on the IGIMF
theory are able to reproduce, among other things, the [$\alpha$/Fe] of
the most metal-poor MW stars, in particular if the slope $\beta$ of
the embedded cluster mass function is equal to 2.  Concerning these
results, it is important to remark that they are based on the WW95
yields\footnote{Actually, they are based on the Fran{\c c}ois et al.
  (2004) yields, but the difference between Fran{\c c}ois's yields and
  WW95 ones is negligible for the chemical elements we are analysing.}
who, however, calculated nucleosynthetic stellar products only for
stars in the interval of initial masses [12,40] M$_\odot$ ([11,40]
M$_\odot$ if the initial composition is solar).  Below 12 M$_\odot$,
yields are usually interpolated (and they are extrapolated above 40
M$_\odot$).

Such an interpolation is expected to have measurable effects on the
computed abundances, since for a standard IMF, a non-negligible mass
fraction of stars forms in the mass range between 8 M$_\odot$ and 12
M$_\odot$.  In this Section, we aim at testing the effects regarding
the assumption of the yields in this mass range, and assess how this
affects the computed abundance ratios.

Assuming that all stars with initial masses larger than 8 M$_\odot$
end their lifetimes as SNeII (i.e. assuming $m_{thr}=8$), by means of
Eq. \ref{eq:plateau} we can explore the effect of the yields in the
interval of initial masses [8,12] M$_\odot$ on the [$\alpha$/Fe] MW
plateau.  The IGIMFs are calculated as in R09, using SFR=0.5 M$_\odot$
yr$^{-1}$ and taking as reference value $\beta=2$ (see Table
\ref{table1}).  We take as the reference set of WW95 yields the one
with 10$^{-4}$ Z$_\odot$ metallicity, case B.  Gibson (1998) showed
that the oxygen yields do not significantly depend on metallicity, at
least for initial metallicities larger than 0.

\begin{figure*}
\centering
\begin{tabular}{cc}
\includegraphics[width=4.5cm, angle=270]{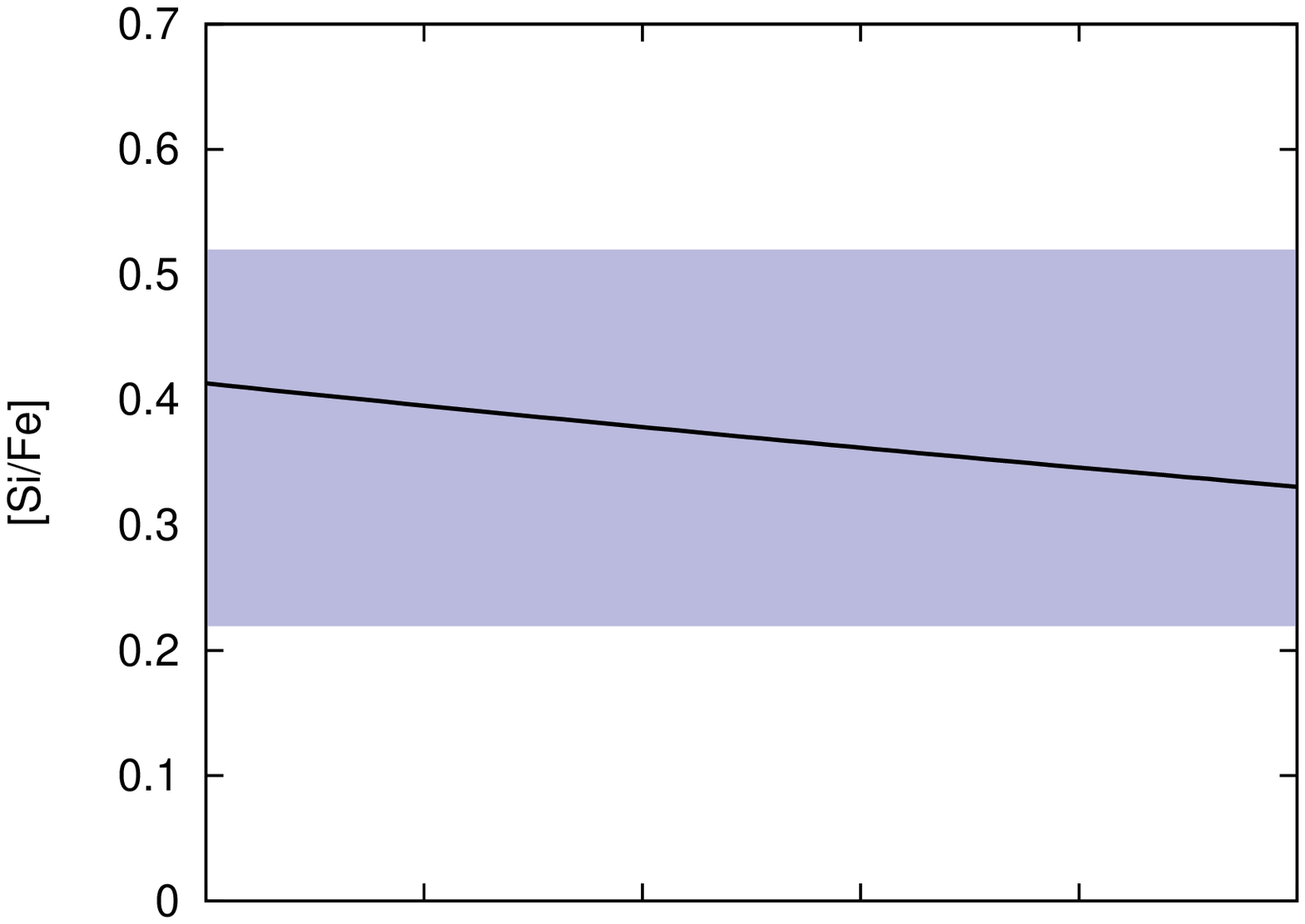} &
\includegraphics[width=4.5cm, angle=270]{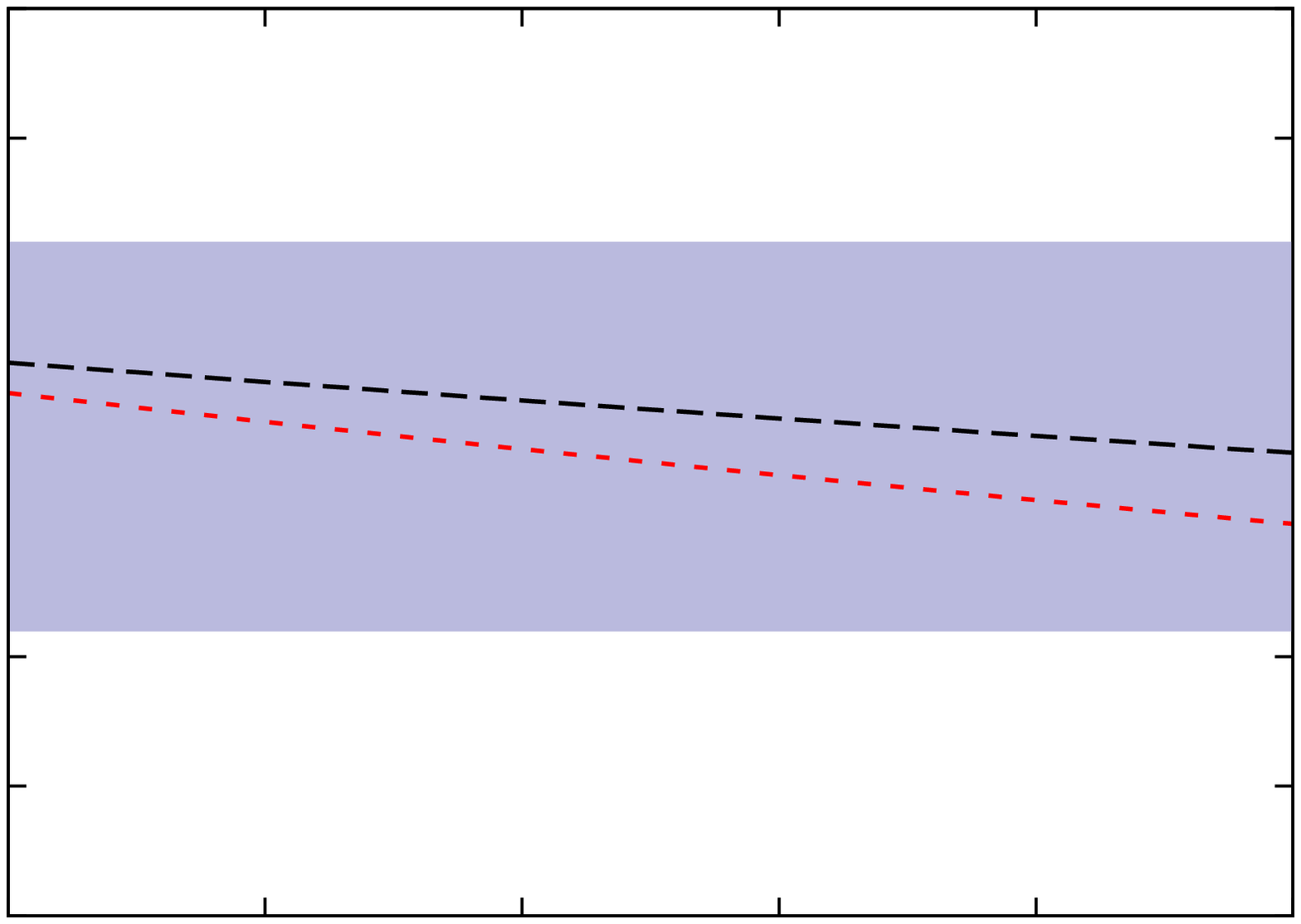}\\
\includegraphics[width=4.5cm, angle=270]{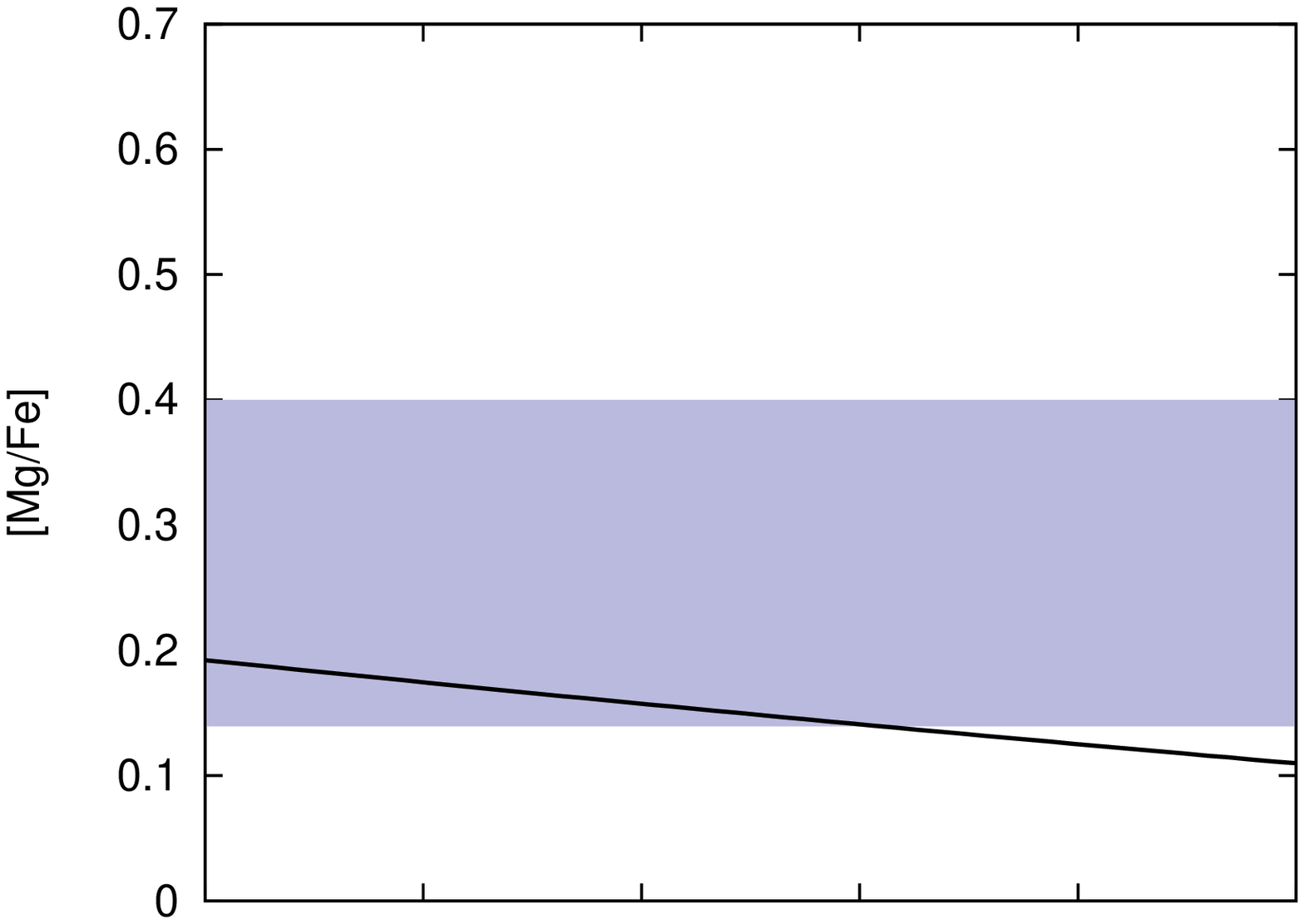} &
\includegraphics[width=4.5cm, angle=270]{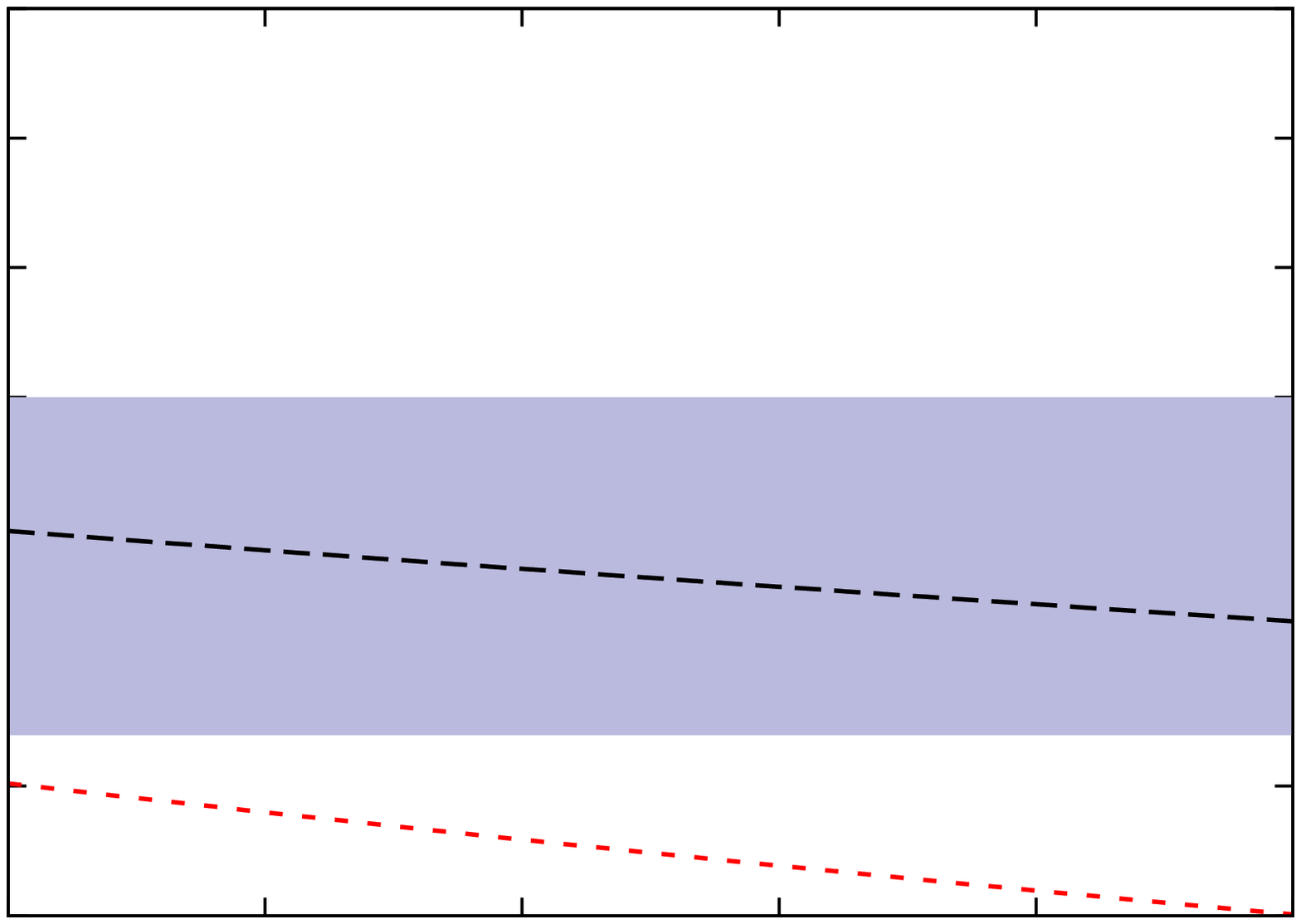} \\
\includegraphics[width=4.5cm, angle=270]{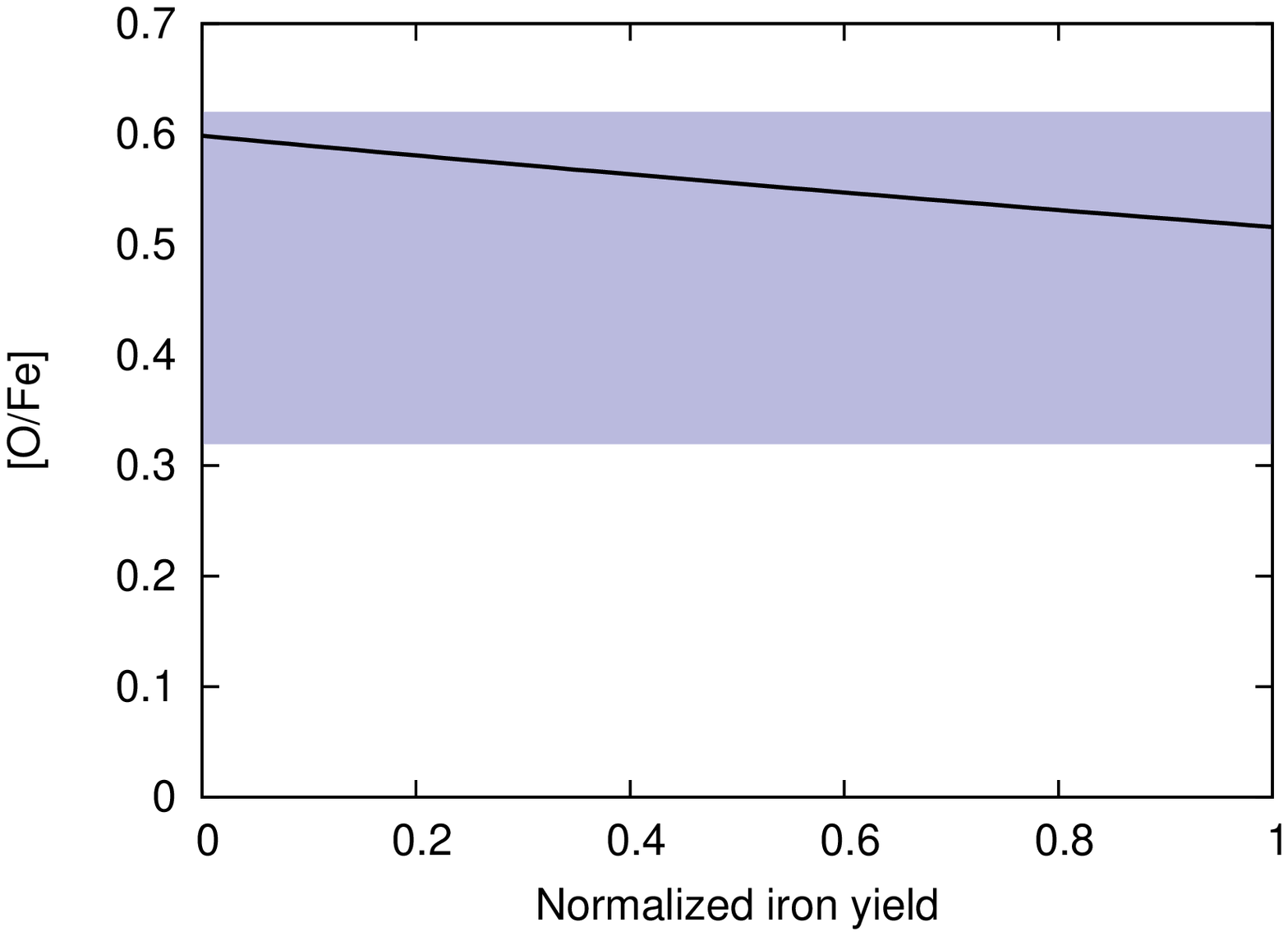}&
\includegraphics[width=4.5cm, angle=270]{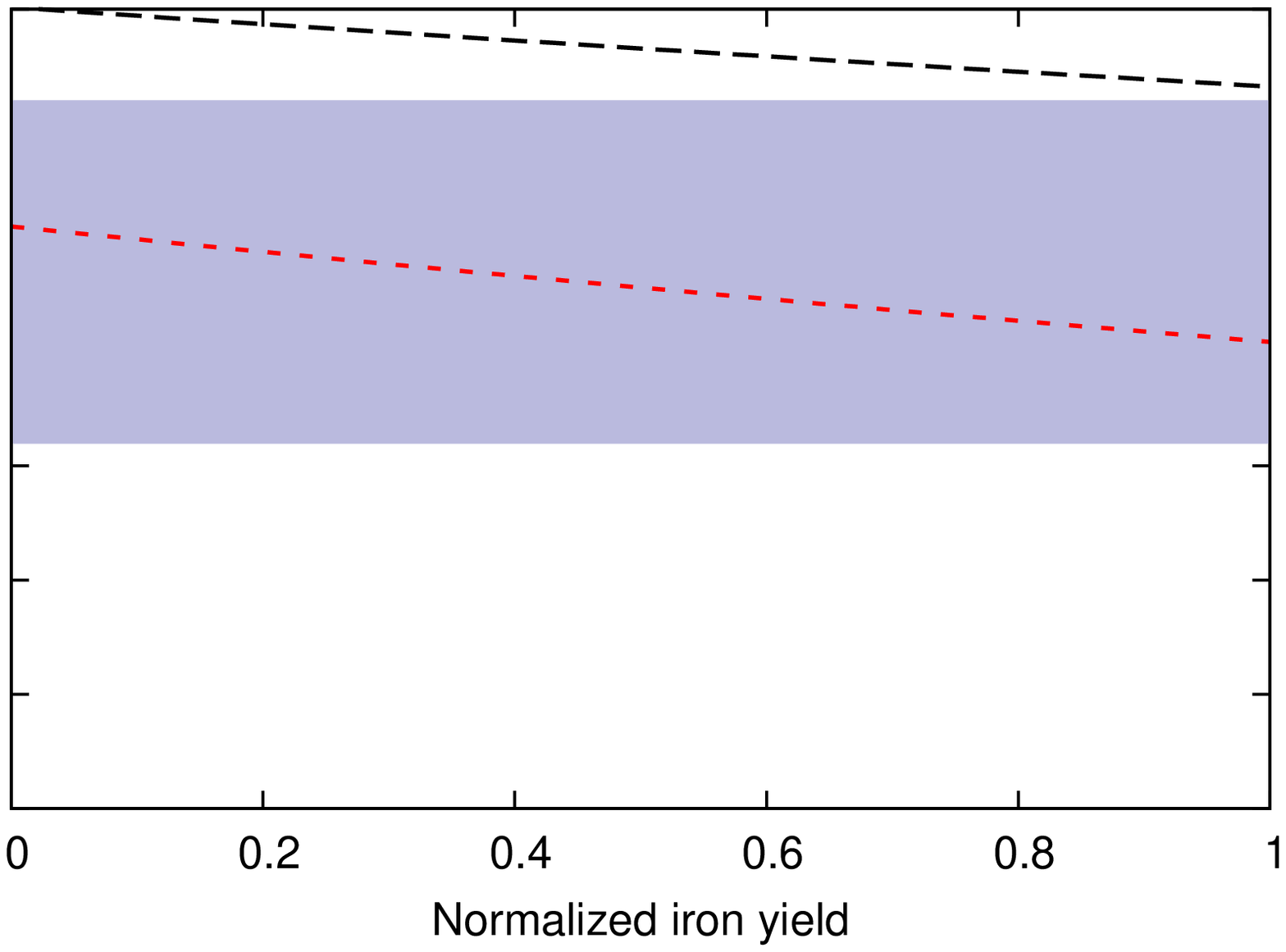} 
\end{tabular}
\caption[]{\label{fig:fit2} Left panels: [O/Fe] (lower panel), [Mg/Fe]
  (middle panel) and [Si/Fe] (upper panel) obtained applying eq.
  \ref{eq:plateau} with $m_{thr}=$ 8 M$_\odot$, with a variable Fe
  yield at 10 M$_\odot$, expressed through the normalised value ${\hat
    y}_{Fe}$ (see eq.  \ref{eq:normfeyield}).  IGIMF parameters are
  the ones reported in Table \ref{table1}.  Shaded areas in each panel
  are the average MW [$\alpha$/Fe] plateau values as obtained from
  Cayrel et al. (2004) $\pm 1 \sigma$. Right panels: Same as left
  panels but with different values of $\beta$: $\beta=1.00$
  (long-dashed lines) and $\beta=2.35$ (short-dashed lines).}
\end{figure*}

Of course what determines the final [$\alpha$/Fe] is the
ratio between $\alpha$-element yields and Fe yields in this interval,
not their absolute value.  Therefore, we fix the Si, Mg and O yields
at 10 M$_\odot$ as the average between the 12 M$_\odot$ yields and the
8 M$_\odot$ yields calculated by van den Hoek \& Groenewegen (1997)
and leave the 10 M$_\odot$ Fe yield $y_{Fe, 10}$ as a free parameter,
which is allowed to vary between the 8 M$_\odot$ yield ($y_{Fe, low}$)
predicted by van den Hoek \& Groenewegen (1997) and the 12 M$_\odot$
yield ($y_{Fe, up}$) of WW95.  The yields in the intervals ]8,10[ and
]10,12[ M$_\odot$ are linearly interpolated between the tabulated
values.  We define the ``normalised'' iron yield
\begin{equation}
\label{eq:normfeyield}
{\hat y}_{Fe}=\frac{y_{Fe, 10}-y_{Fe, low}}{y_{Fe, up}-y_{Fe, low}}.
\end{equation}
\noindent 
The quantity ${\hat y}_{Fe}$ varies between 0 and 1.  The [O/Fe],
[Mg/Fe] and [Si/Fe] predicted by means of Eq.  \ref{eq:eqgibson} as a
function of ${\hat y}_{Fe}$ are shown in Fig.  \ref{fig:fit2} (left
panels).  The IGIMF parameters assumed to calculate these
[$\alpha$/Fe] ratios are the reference values reported in Table
\ref{table1}.  In particular, $\beta=2$ is assumed.  These values are
in agreement with the results of Calura et al.  (see their Fig. 3,
upper panel, for [O/Fe], Figs. 6, 9 and 14, lower panels, for
[Si/Fe])\footnote{It is worth pointing out that C10 adopted the
  Asplund et al.  (2004) solar abundances; however these values differ
  very little from our assumptions, so the comparison with Calura et
  al.'s results is still meaningful.} and are within the observed
ranges, irrespective of ${\hat y}_{Fe}$, with the exception of [Mg/Fe]
for ${\hat y}_{Fe}>0.6$.\footnote{Notice that here we have assumed the
  photospheric solar Mg abundance of Asplund et al. (2009).  Had we
  chosen the meteoric Asplund et al.'s abundance 12+log(Mg/H)=7.53,
  the [Mg/Fe] would be consistent with the observed Milky Way plateau
  for the whole interval in ${\hat y}_{Fe}$.}

From Fig. \ref{fig:fit2} we can observe that the dependence of the
[$\alpha$/Fe] ratio on ${\hat y}_{Fe}$ is moderate, with variations of
$\sim$ 0.08 dex at most.  This dependence, together with the changes
many other parameters produce in the [$\alpha$/Fe] plateaus, is
summarised in Table \ref{table3}.  From Fig.  \ref{fig:fit2}, left
panels, we can also observe that the best match between observations
and model results is obtained for ${\hat y}_{Fe}$ between 0 and 0.5.
We are mostly interested in the [Mg/Fe] abundance ratios, thus we
choose ${\hat y}_{Fe}=0$ as our reference value (see
Tab.~\ref{table2}), since it allows us to reproduce at best the
[Mg/Fe] plateau as observed in local stars.
  
\begin{table*}
  \caption{Modified parameters (as compared to the reference values 
    of Tables \ref{table1} and \ref{table2}) and their effect on the 
    predicted [$\alpha$/Fe] ratios.} 
\label{table3}      
\centering                          
\begin{tabular}{c c c c }        
\hline\hline                 
Quantity  &  Modified value/   & [$\alpha$/Fe] variation (dex) & Sect. \\ 
  &  set of values   & & \\ 
\hline
 ${\hat y}_{Fe}$ & [0,1] & 0.08 & \ref{sec:resmw}, \ref{sec:resgal}  \\
 $m_{thr}$ & 10 M$_\odot$ & 0.04 & \ref{sec:app} \\
 $\beta$ & 1; 2.35 & 0.2 & \ref{sec:resmw} \\
 $y_{Fe, 150}/y_{Fe, 40}$ & [1,10$^4$] & 0.005 & \ref{sec:app} \\
 $<SFR>_{1 Gyr}$ & 100 M$_\odot$ yr$^{-1}$ & 0.015 & \ref{sec:app} \\
 $M_{\rm ecl, min}$ & 20 M$_\odot$ yr$^{-1}$ & $<$0.01 & \ref{sec:app} \\
 $m_{\rm max, *}$ & 300 M$_\odot$ yr$^{-1}$ & 0.03--0.04 & \ref{sec:app} \\
 $\alpha_2$ & 2.3 + 0.0572 [Fe/H] & 0.2 & \ref{sec:disc}\\
 (Mild Z dependence) & & & \\
 $\alpha_1$, $\alpha_2$ & $\alpha_1=$ 1.3+0.5 [Fe/H] &0.7--0.8 dex 
& \ref{sec:disc}\\
 (Hard Z dependence) & $\alpha_2=$ Min(2.63+0.66[Fe/H],2.3) & & \\
\hline                                   
\end{tabular}
\end{table*}

\begin{figure}
\centering
\includegraphics[width=9cm]{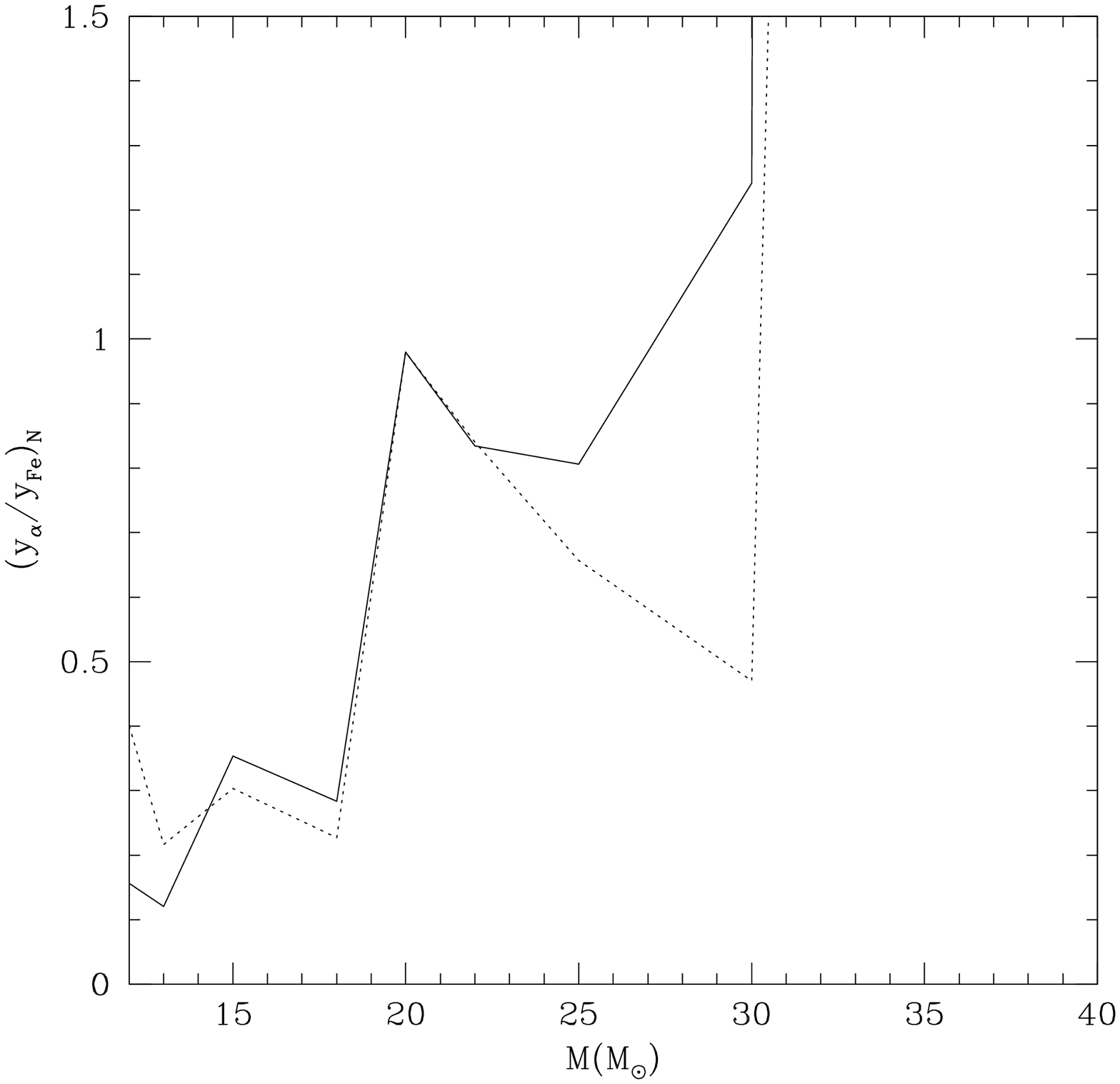}
\caption[]{\label{fig:yieldratio} $y_O/y_{Fe}$ (solid line) and 
$y_{Si}/y_{Fe}$ (dotted line) tabulated by WW95 for Z=10$^{-4}$ 
Z$_\odot$, as a function of the initial stellar mass.  The yield 
ratios have been normalised to the corresponding yield ratio for 
a 20 M$_\odot$ star. }
\end{figure}

Fig. \ref{fig:fit2} (right panels) shows the dependence of the
[$\alpha$/Fe] Milky Way plateaus on $\beta$.  It is clear that $\beta$
affects quite significantly the results (see also R09 and C10).  The
fact that [O/Fe] with $\beta=1$ and [Mg/Fe] with $\beta=2.35$ do not
fit the observations corroborates our choice of $\beta=2$ as reference
value.  We can also observe that the [Si/Fe] results are less
dependent on $\beta$ than the [O/Fe] and [Mg/Fe] ratios, in agreement
with the results of C10.  This is due to the fact that the pattern of
the Fe yields as a function of initial stellar mass is more similar to
the Si yields pattern as compared to the oxygen and magnesium one.  In
other words, the mass dependence of the ratio $y_{Si}/y_{Fe}$ is
flatter than the function that describes $y_O/y_{Fe}$ (or
$y_{Mg}/y_{Fe}$) as a function of mass.  This is shown in Fig.
\ref{fig:yieldratio}, where the yield ratios have been divided by the
ratios of stellar yields at 20 M$_\odot$.  From this figure it is also
worth noticing the large Si yield at 12 M$_\odot$, which has
consequences for our results.

It is important to explain in more detail the reason of our choice of
an initial metallicity of Z=10$^{-4}$ Z$_\odot$ (see also Sect.
\ref{sec:calc}).  The WW95 set of yields at 10$^{-2}$ Z$_\odot$ can
not be considered because the MW plateau extends to metallicities much
below this value.  The most metal-poor stars in the Cayrel et al. 2004
sample have in fact [Fe/H] slightly larger than -4.  This forces us to
consider only initial metallicities equal or smaller than 10$^{-4}$
Z$_\odot$.  We have considered the predicted WW95 yields at Z=0 (case
B).  Metal-free stars produce a limited amount of $\alpha$-elements,
resulting in very low [$\alpha$/Fe] ratios.  Only the models with
$\beta=1$ attain [$\alpha$/Fe] larger than 0.2 dex and none of the
considered models reach the assumed [$\alpha$/Fe] MW plateau value.
Even the more recent, detailed calculations of nucleosynthetic yields
from metal-free massive stars by Heger \& Woosley (2010) do not agree
with the Cayrel plateaus for many elements (see their Fig. 12).  We
thus discard this set of yields.  Since the WW95 set of yields does
not consider metallicities intermediate between Z=10$^{-4}$ Z$_\odot$
and Z=0, it is clear that in the present study the only possible
choice for the initial metallicity is Z=10$^{-4}$ Z$_\odot$.

We have thus seen in this section that, besides the SFR (the
independent variable in our study), our calculations of the
[$\alpha$/Fe] plateaus depend sensitively on the choice of $\beta$
and, to a lesser extent, on the choice of ${\hat y}_{Fe}$, i.e. on the
choice of the yields in the interval [8,12] M$_\odot$.  Many other
parameters have been investigated in the Appendix and they turn out to
have a negligible effect on our results (see Table \ref{table3}).

\subsection{The [$\alpha$/Fe] ratios of the most metal-poor stars in 
dwarf galaxies}
\label{sec:resgal}

\begin{figure}
\centering
\includegraphics[width=9cm]{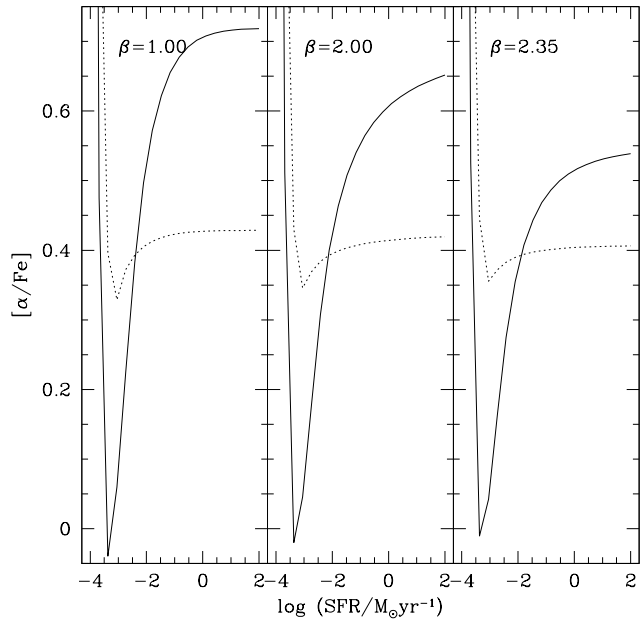}
\caption[]{\label{fig:alphafe2} [O/Fe] (solid lines) and [Si/Fe]
  (dotted lines) ratios as a function of the galaxy SFR, for different
  values of $\beta$ (indicated on the top of each panel).  We assume
  in this plot the reference value ${\hat y}_{Fe}=0$.}
\end{figure}

We can now relax the hypothesis of a fixed SFR and explore the
predicted values of the [$\alpha$/Fe] plateau in dwarf galaxies,
characterised by mild SFRs.  In order to illustrate our results, we
plot in Fig.  \ref{fig:alphafe2} the calculated [O/Fe] and [Si/Fe]
plateaus as a function of SFR in the case in which ${\hat y}_{Fe}=0$
and $m_{thr}=8$ M$_\odot$ (i.e.  our reference values).  Although our
reference value of $\beta$ is 2, we show in this plot also the results
obtained with different values of this parameter.  

As one can see from this figure, [O/Fe] shows very large variations as
a function of SFR below SFR $\simeq$ 0.1 M$_\odot$ yr$^{-1}$.  Again,
the variation of [Si/Fe] is more limited (at least for SFRs larger
than 10$^{-3}$ M$_\odot$ yr$^{-1}$) and it is significant only below
SFR $\simeq$ 10$^{-1}$ M$_\odot$ yr$^{-1}$.  The main reason of this
different behaviour is still the different pattern of $y_O/y_{Fe}$ and
$y_{Si}/y_{Fe}$ shown in Fig.  \ref{fig:yieldratio}.  From Fig.
\ref{fig:alphafe2} we can notice that the [$\alpha$/Fe] of the models
with the weakest SFR are even higher than the MW plateau
[$\alpha$/Fe].  This is due to the fact that, below 12 M$_\odot$, the
Fe production is negligible.  It is only due to secondary production
which, being the considered metallicity of the stellar population
10$^{-4}$ Z$_\odot$, is very tiny.  Also the O and Si production is
very low, but still much larger than the Fe production.  However, due
to the very limited quantity of Fe and $\alpha$-elements synthesised
by these model galaxies, these results cannot be safely compared to
the observations.  

The [O/Fe] has a pronounced minimum at log SFR $\simeq$ -3.7.  The
difference between this minimum and the [O/Fe] MW plateau is of the
order of $\simeq$ 0.5 dex.  Overall, if we neglect the initial steep
decline of [O/Fe], there are variations of the order of up to 0.7 dex
of the [O/Fe] ratios as a function of the SFR.  If we repeat this
calculation with ${\hat y}_{Fe}=0.5$ (which is still compatible with
the Milky Way [$\alpha$/Fe] plateau) the resulting [O/Fe] ratios reach
minima of the order of $\sim -0.2$ dex, almost 0.7 dex below the MW
plateau value.  Overall, the differences with the model with ${\hat
  y}_{Fe}=0$ are of the order of $\sim$ 0.1 dex, at least for SFR $>$
10$^{-3}$ M$_\odot$ y$^{-1}$.

A general conclusion we can draw from Fig. \ref{fig:alphafe2} is that,
irrespective of the SFR, the predicted [Si/Fe] plateau of dwarf
galaxies is always $\geq 0$ and never departs significantly from the
observed MW plateau.  On the other hand, the predicted [O/Fe] plateau
can differ by up to 0.5 dex compared to the MW halo value.  Concerning
oxygen abundances, one might argue that they are known to be
relatively low in a large fraction of the red giants within Galactic
globular clusters (e.g. Carretta et al. 2009), and that this could be
true also in dwarf galaxies.  However, this point should not obviate
any of our analysis, since O-poor stars in GC are likely
second-generation stars born out of gas containing the ejecta of
older, first generation stars, as indicated also by the
well-established O/Na anti-correlation in GCs (Gratton et al. 2004;
Gratton et al. 2012).  Moreover, as suggested by Carretta et al.
(2010), such an anti-correlation is not observed in the body of
Saggitarius field stars.  The same may be true for other dwarf
galaxies. At the present stage, in the light of the results of
Carretta et al. (2010), the [O/Fe] values adopted here for the stars
of local dwarfs galaxies can be considered as representative of the
bulk of their stellar population.

Although illustrative, the [O/Fe] and [Si/Fe] ratios can not be easily
compared with available observations.  The determination of [O/Fe]
ratios in DGSs is difficult and it has been performed only by a
handful of authors.  There are more data available about [Si/Fe] but,
as can be seen from Fig.  \ref{fig:alphafe2}, we predict moderate
variations of [Si/Fe] as a function of the SFR, hence it is difficult
from this abundance ratio to constrain the IMF or the SFR.  A larger
amount of data is available for the [Mg/Fe] abundance ratios (see also
Sect.  \ref{sec:afe_obs}).  In Table \ref{table4} we collect available
data of [Mg/Fe] ratios in stars of the DGSs Draco, Carina, Sextans and
Sculptor, having [Fe/H]$<-2$.  Average [Mg/Fe] ratios for each galaxy
have been calculated by considering data coming from different papers
(reported in the Table, as well). Since the average [Mg/Fe] in Cayrel
et al.  (2004) is $\simeq 0.27$, we can notice from Table \ref{table4}
that average [Mg/Fe] ratios in the four considered dwarf galaxies lie
below this MW plateau value, although the differences are of the order
of $\sim$ 0.15 dex at most.

We choose to concentrate on the range of iron abundances with
[Fe/H]$<-2$ in order to be consistent with the range of [Fe/H] ratios
of the Cayrel (2004) sample for MW halo stars.  However, it is
important to remind the reader that some metal-poor stars with
relatively low [Mg/Fe] ratios have [Fe/H] only slightly below -2.  If
we consider as a threshold metallicity [Fe/H]=-2.3 instead of -2, the
average [Mg/Fe] ratios in Draco and Sculptor remain unaltered, the
average [Mg/Fe] in Sextans increases by only $\sim$ 0.02 dex, whereas
the average [Mg/Fe] ratio in Carina changes dramatically from 0.144 to
0.328.  However, looking at Starkenburg et al. (2013) one could also
argue that extremely metal-poor stars (i.e. stars with [Fe/H]$<$-3)
have also [Mg/Fe] ratios below the average and that the [Mg/Fe] ratio
has a peak at [Fe/H]$\simeq$-3.
\begin{table*}
  \caption{Average [Mg/Fe] ratios with relative errors for the most 
    metal-poor stars (stars with 
    [Fe/H] below -2) in the Draco, Carina, Sextans and Sculptor DGSs.} 
\label{table4}      
\begin{centering}                          
\begin{tabular}{c c c c c}        
  \hline\hline                 
  Galaxy  &  Average [Mg/Fe] & $\sigma$ & References & Allowed log 
  SFRs (in M$_\odot$ y$^{-1}$)\\ 
  \hline
  Draco & 0.228 & 0.051 & Shetrone et al. (2001) & log SFR$>$-0.52 \\
  & & & Cohen \& Huang (2009)$^{(1)}$\\
  Carina & 0.144 & 0.099 & Koch et al. (2008) & -1.82 $<$ log SFR 
  $<$ 1.70\\
  & & & Venn et al. (2012)\\
  & & & Lemasle et al. (2012)\\
  Sextans & 0.166 & 0.029 & Tafelmeyer et al. (2010) & -1.10 $<$ log SFR 
  $<$ -0.15 \\
  & & & Shetrone et al. (2001)\\
  & & & Aoki et al. (2009)\\
  Sculptor & 0.162 & 0.055 & Starkenburg et al. (2013) & -1.40 $<$ 
  log SFR $<$ 0.45\\
  & & & Tafelmeyer et al. (2010)\\
  \hline 
\end{tabular}
\end{centering}

\noindent
$^{(1)}$ Notice that 
Cohen \& Huang (2009) had to increase the 
[Mg/Fe] determined by Cayrel et al. (2004) by 0.15 dex to make 
it compatible with their measurements.
\end{table*}

\begin{figure*}
\centering
\begin{tabular}{cc}
\includegraphics[width=5.5cm, angle=270]{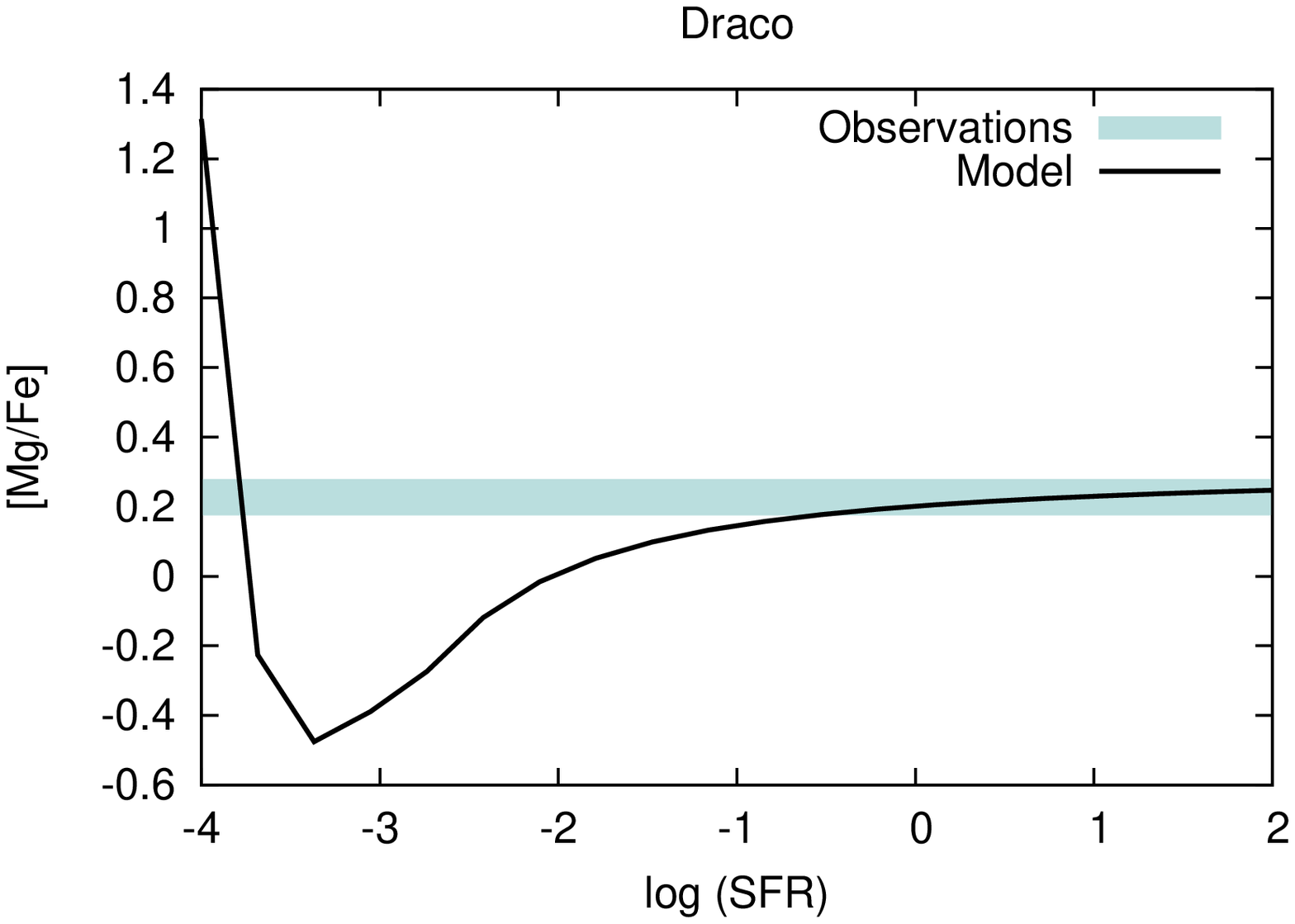}&
\includegraphics[width=5.5cm, angle=270]{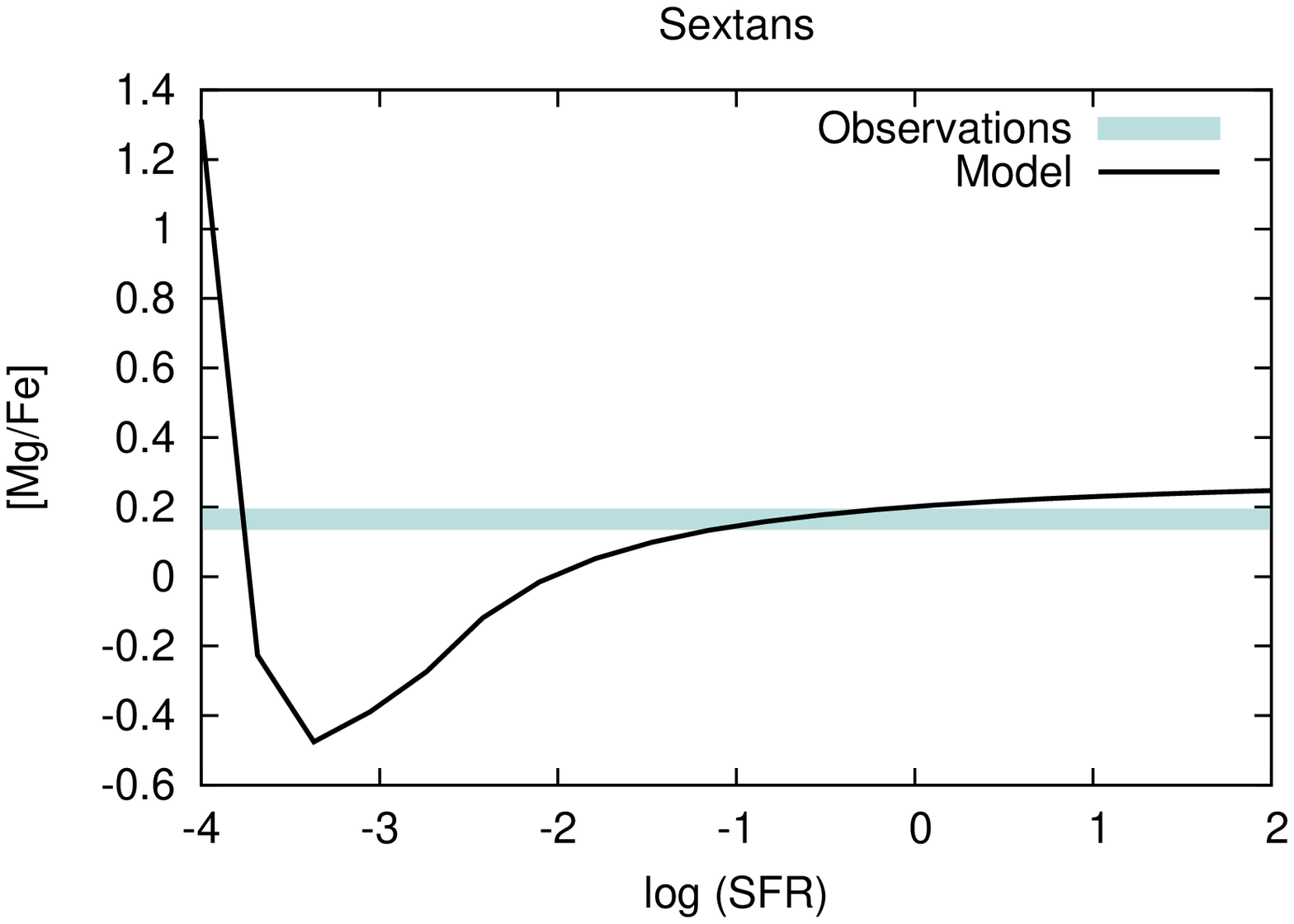} \\
\includegraphics[width=5.5cm, angle=270]{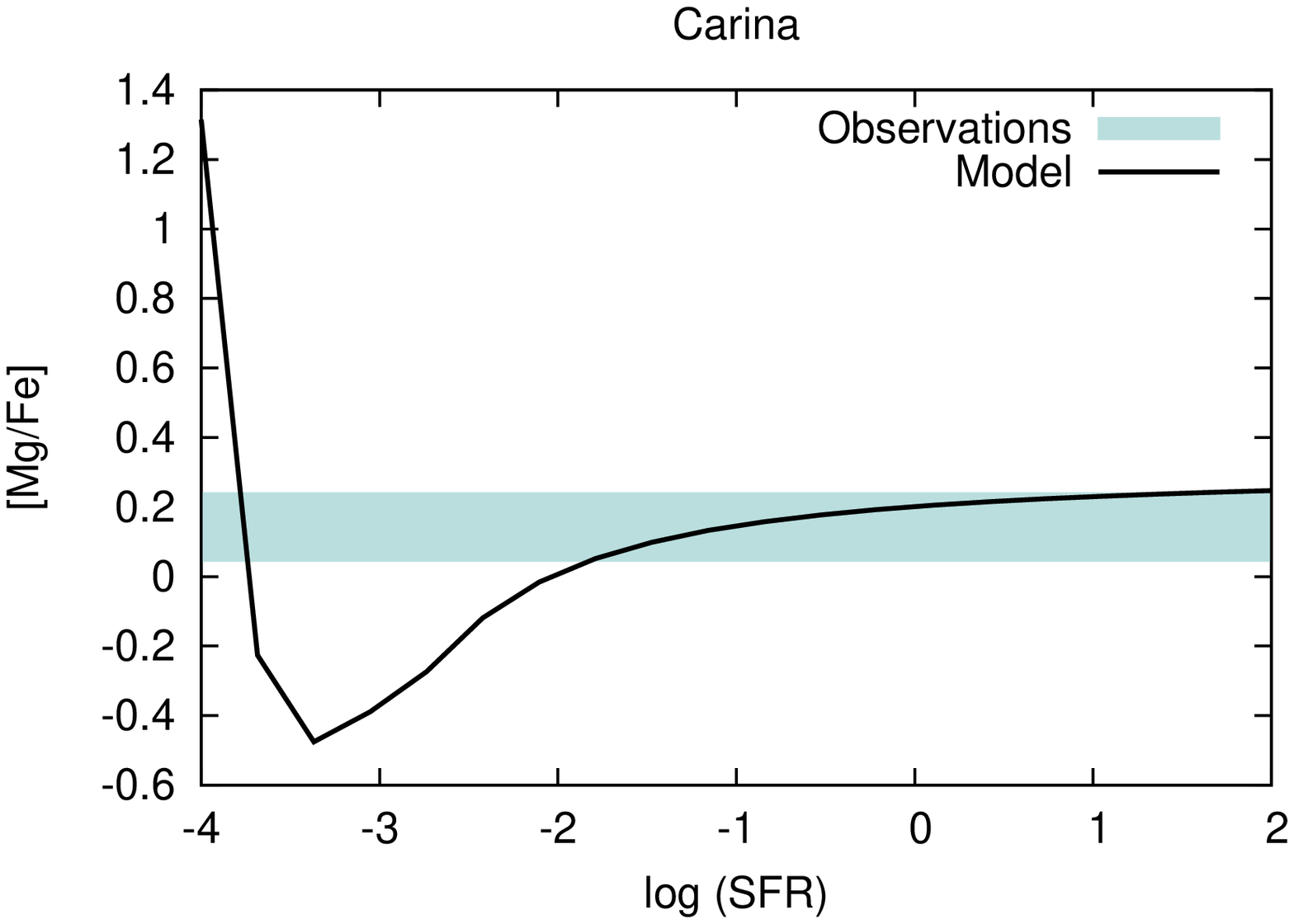}&
\includegraphics[width=5.5cm, angle=270]{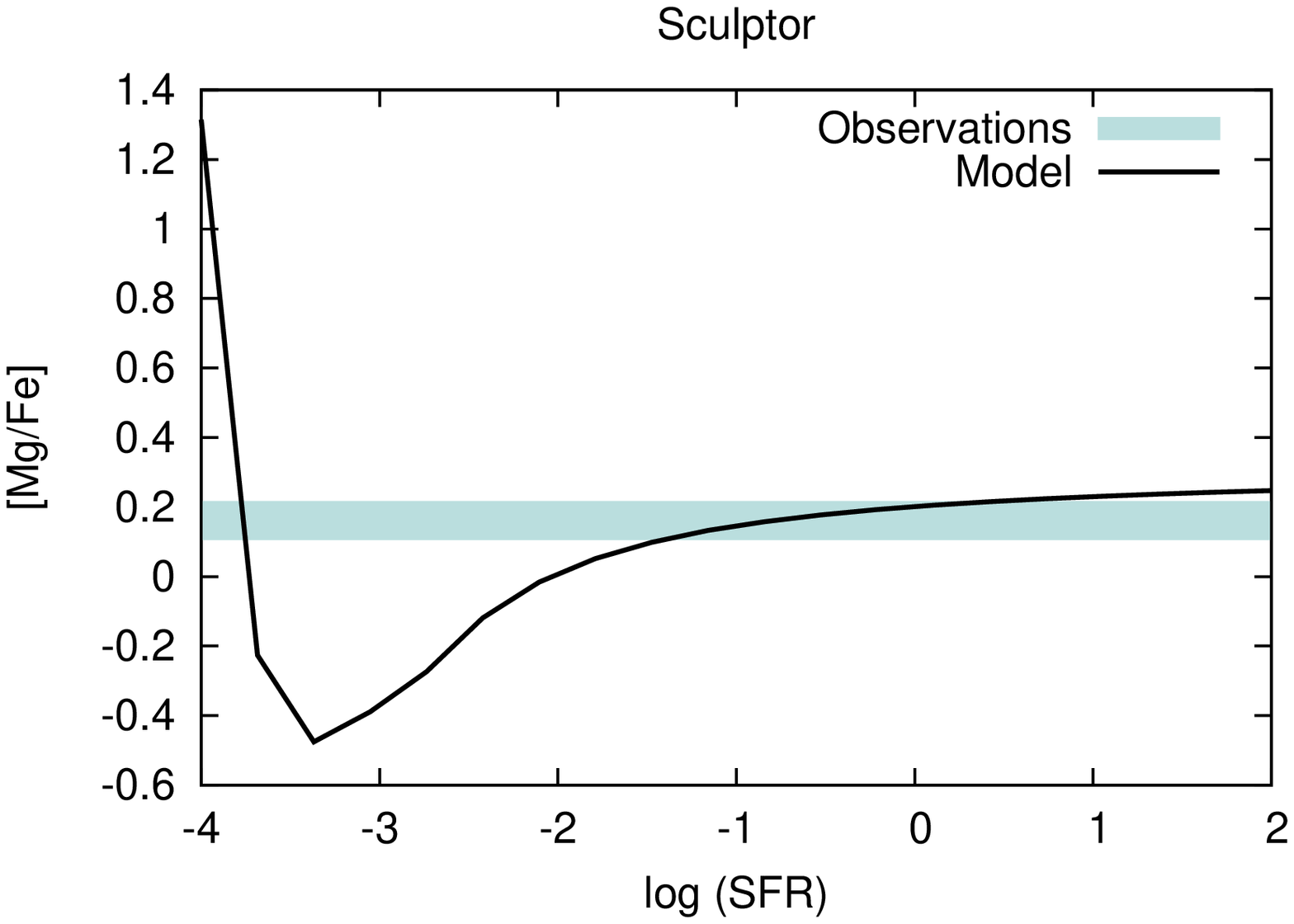} \\
\end{tabular}
\caption[]{\label{fig:comparison} Comparison of computed [Mg/Fe]
  plateau as a function of the SFR (solid lines) with observed
  averages of the most metal-poor stars in four well-studied DGSs:
  Draco, Carina, Sextans and Sculptor (shaded areas, see Table
  \ref{table4}).}
\end{figure*}

We calculate the variation of the [Mg/Fe] abundance ratio as a
function of the SFR exactly as we did for the [O/Fe] and [Si/Fe]
abundance ratios.  We concentrate in what follows on the reference
values of ${\hat y}_{Fe}=0$ and $\beta=2$.  We compare the results of
our models with the data presented in Table \ref{table4} for the dwarf
galaxies Draco, Sextans, Carina and Sculptor in Fig.
\ref{fig:comparison}.  Given the uncertainties in the determinations
of the SFR in the earliest phases of galaxy evolution described in
Sect. \ref{sec:sfh_obs}, we draw the measurements of the observed
[Mg/Fe] ratios as horizontal shaded areas.

At least in principle, this plot could be used to constrain the early
SFR in the four considered dwarf galaxies, in the framework of the
IGIMF theory.  From this plot it seems that the early SFR of these
galaxies is either very low, with SFRs below 10$^{-3}$ M$_\odot$
yr$^{-1}$, or moderately intense, with SFRs larger than 10$^{-2}$
M$_\odot$ yr$^{-1}$, depending on the galaxy.  The former possibility
should be ruled out because of the large variations of the [Mg/Fe] as
a function of the SFR, which would make unlikely the observations of
similar (within a factor of 0.3 dex) [Mg/Fe] ratios in different
galaxies.  The latter is a viable possibility and we have summarised
the range of allowed SFRs for each galaxy in Table \ref{table4}.
However, one should not forget the uncertainties implied by this kind
of comparison, among which: $(i)$ not always abundance ratios are
determined with the same methodologies by different groups and
different methodologies can lead to different results.  $(ii)$ Even
the solar abundance of Mg is still uncertain.  $(iii)$ At variance
with the MW halo, the statistics of very metal-poor stars in DGSs is
still quite poor.  $(iv)$ There are uncertainties related to the
parameters and to the formalism of the IGIMF theory.  Some of these
uncertainties have been discussed in this paper and summarised in
Table \ref{table3}.

\section{Discussion and results with a metallicity dependent IGIMF}
\label{sec:disc}

\subsection{Results with a metallicity-dependent IGIMF}
\label{sec:zdepigimf}
As already mentioned, it has been recently suggested that metallicity
can play a role in shaping the IMF in galaxies.  In particular,
galaxies with low metallicities should have top-heavier IMFs than high
metallicity galaxies with the same SF rates (see 
M12, Kroupa et al. 2013 and references therein).  It is important to
stress that the results of M12 are mostly based on studies of the mass
distribution in globular clusters, therefore it is not clear whether
they can be directly applied to galaxies.  Nevertheless, we believe
that it is interesting to study the [$\alpha$/Fe] ratios in the
framework of this new IGIMF formulation also because it is the first
time to our knowledge that the chemical composition and evolution of a
galaxy having a metallicity-dependent IGIMF has been studied.

M12 consider indeed the combined effect of cluster densities and
metallicities.  Since we have no information on the densities of our
model galaxies, we will consider only the results of M12 concerning
the metallicity dependence of the IMF.  If we assume that all the star
clusters in galaxies, irrespective of their masses, always have the
same density, we end up with a mild dependence of the IGIMF on the
metallicity ({\it mild model}).  The metallicity of our model dwarf
galaxies at different masses (and SFRs) is calculated according to eq.
(2) of Mannucci et al. (2010; the so-called {\it fundamental
  metallicity relation}).  The correlation between SFR and mass of a
model galaxy is obtained exactly as in R09.  The obtained (O/H) is
converted into [Fe/H] by means of the solar abundances of Asplund et
al. (2009).\footnote{This calculation of [Fe/H] is quite approximate
  and a more detailed calculation of a metallicity-dependent IGIMF in
  a galaxy of a given mass results from an iterative procedure.  Our
  main aim here is to have a reasonable dependence of the metallicity
  with the SFR (namely, to assign a reasonable metallicity [Fe/H] to
  each model galaxy) and, through it, to compare calculations with and
  without metallicity-dependent IMFs.  Thus, we prefer to use a simple
  estimate of [Fe/H] and hence of the IGIMF}.  The relation we obtain
in the end, adopting eq. (15) of M12 with a constant value of the
cluster density $\rho_{cl}$ is:

\begin{equation}
\label{eq:zdep_mild}
\alpha_2=2.3+0.0572 [Fe/H],
\end{equation}
\noindent
where $\alpha_2$ is the IMF slope for high-mass stars already
introduced in Sect. \ref{sec:igimf} and in Table \ref{table1}.  This
formula ensures that a galaxy like the Milky Way has a present time
high-mass IMF slope of 2.3, in agreement with the assumptions of M12,
and very similar to the high mass IMF slope of 2.35 adopted by R09.
It is important to stress that we modify the IMF {\it within each star
  cluster} according to this formula.  Once a distribution function of
star clusters has been assumed, the integrated IMF is then calculated
by integrating the IMF of the various star clusters, according to the
standard recipes (see Sect.  \ref{sec:igimf} and R09).

\begin{figure}
\centering
\includegraphics[width=6cm,angle=270]{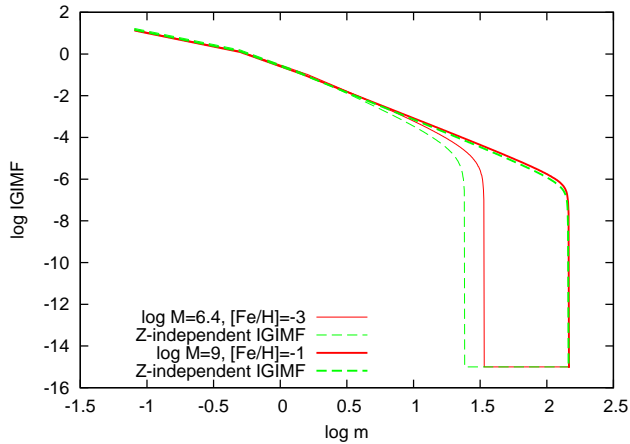}
\caption[]{\label{fig:comp1} The comparison of the logarithm of the
  IGIMF as a function of the stellar mass ($m$) obtained by means of
  the mild model (eq. \ref{eq:zdep_mild}; solid lines) for two
  representative values of the metallicity ([Fe/H]=-3: light line;
  [Fe/H]=-1: heavy line).  The two dashed lines represent calculations
  made for two metallicity-independent IGIMFs (with $\beta=2.00$) at
  two different galaxy masses: log M=6.4 (light line); log M = 9
  (heavy line).}
\end{figure}

Considering for simplicity only the reference case $\beta=2.00$, the
comparison between the standard and the metallicity dependent IGIMF
for two representative metallicities is presented in Fig.
\ref{fig:comp1}.  As one can see from this figure, at [Fe/H]=-1 the
metallicity-dependent IGIMF is very similar to the reference,
Z-independent one (the one obtained in R09).  At [Fe/H]=-3 the
difference is more significant but still not very large.  The effect
of this metallicity-dependent IMF is to make IGIMFs in low-mass (hence
metal-poor) galaxies flatter, thus partially counter-balancing the
effect of the SFR on the IGIMF.  However, since the metallicity effect
is mild in this formulation, dwarf galaxies are still characterised by
steeper IGIMFs than large ones.

\begin{figure}
\centering
\includegraphics[width=6cm,angle=270]{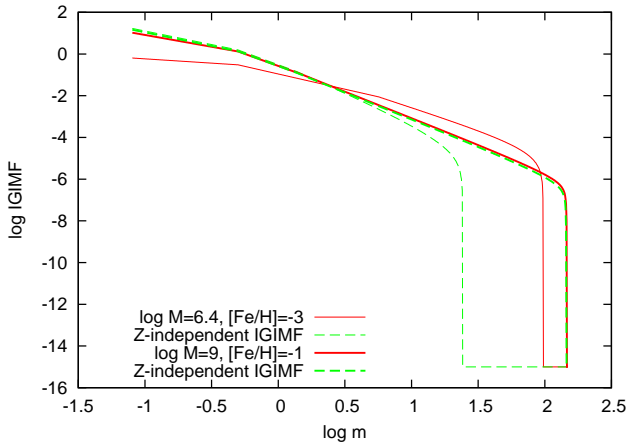}
\caption[]{\label{fig:comp2} As Fig. \ref{fig:comp1} but for the 
hard dependence of the IGIMF on metallicity  (eqs. \ref{eq:zdep_hard_1} 
and \ref{eq:zdep_hard_2})}
\end{figure}

A more pronounced metallicity dependence ({\it hard model}) is
obtained by assuming that also the IMF slope $\alpha_1$ below 0.5
M$_\odot$ in each star cluster depends on [Fe/H].  In agreement with
eq. (12) of M12 we assume:
\begin{equation}
\label{eq:zdep_hard_1}
\alpha_1=1.3+0.5[Fe/H]
\end{equation}
\noindent
Moreover, following eq. (11) and Table 3 of M12 we assume
\begin{equation}
\label{eq:zdep_hard_2}
\alpha_2=Min(2.63+0.66[Fe/H],2.3)
\end{equation}
\noindent
In this way, $\alpha_2$ grows linearly with [Fe/H] until [Fe/H]=-0.5
and then it remains constant at $\alpha_2=2.3$ for [Fe/H]$>$-0.5.  The
comparison between the standard and the metallicity dependent IGIMF in
the hard model for two representative metallicities is presented in
Fig.  \ref{fig:comp2}.  As one can see from this figure, at [Fe/H]=-1
the metallicity-dependent IGIMF is still quite similar to the
reference one but at [Fe/H]=-3 the difference is extremely large.
According to this model, low-mass galaxies, in spite of their low
SFRs, have {\it flatter} IGIMFs than high mass ones.  It is worth
noticing that Zhang et al. (2007) found indeed a dependence of the IMF
slope on the metallicity of the parent galaxy.  Galaxies at higher
metallicities were found to have steeper IMFs, with the slope index
ranging from $\sim$ 1.00 for the lowest metallicity, to $\sim$ 3.30
for the highest metallicity.  Such direct determinations of the IMF
slopes in galaxies are extremely useful to understand how the IMF is
related to the galactic metal content and to assess the validity of
the metallicity-dependent IGIMF put forward by M12.

We should pause here to consider more carefully the assumptions
implied by our semi-analytical calculations.  The fundamental
metallicity relation of Mannucci et al. (2010) does not consider the
detailed chemical evolution of a galaxy.  It just reveals a
correlation between the SFR and the metallicity of a star-forming
galaxy at the moment at which we observe it.  A galaxy with a large
SFR appears to have also a large metallicity.  As we have seen, such a
large metallicity tends to produce steep IMF slopes.  A large SFR on
the other hand tends to make the IMF flatter.  According to the hard
model, the metallicity effect prevails.  Consequently, at the time of
observation, the IMF in high-SFR, high-metallicity galaxies is steeper
than the corresponding IMF in smaller galaxies.

Of course, also the halo of the MW has experienced phases of its
evolution characterised by very low metallicities, during which the
IMF was flat.  To keep track of this time evolution of the IMF, we
need a theoretical approach like the one employed in C10.  However,
our semi-analytical approach still can give us important indications
on the average IMF during the period of the halo formation.  According
to one of the most fundamental results of galactic chemical evolution,
the larger the SFR, the higher the [Fe/H] reached before Type Ia SNe
contribute significantly to the chemical enrichment (Matteucci 2001).
During the formation of the halo, the SFR was relatively high.  That
allowed the [Fe/H] to reach a value of -1, without significant
contribution from Type Ia SNe.  In the framework of the
metallicity-dependent IGIMF, we thus assume that the stars in the MW
halo stem from stellar populations having different metallicities, up
to [Fe/H]=-1.  In the most metal-poor stellar populations the IMF is
flat but it steepens as the metallicity increases.  Plateau stars are
stars for which the contribution from Type Ia SNe is still negligible.
Since the MW plateau stars have metallicities up to [Fe/H]=-1, we
expect IMFs corresponding to metallicities up to that value to be
relevant for the stars we consider in this work.  This is also the
reason why we plot in Figs.  \ref{fig:comp1} and \ref{fig:comp2} the
IMFs of model galaxies with [Fe/H]=-1.  On the other hand, in dwarf
galaxies, characterised by lower SFRs, the iron enrichment by SNeII is
less pronounced and SNeIa start contributing to the chemical evolution
when [Fe/H] is still quite low.  In other words, the plateau extends
up to lower metallicities.  For these galaxies, the IMF corresponding
to [Fe/H]=-1 plotted in Figs.  \ref{fig:comp1} and \ref{fig:comp2}
does not play any role in the chemical evolution of the most
metal-poor, plateau stars of their respective galaxies.  According to
the hard model, we thus expect plateau stars in dwarf galaxies to be
made of stellar populations with low metallicities, characterised by
flat IMFs as shown in Fig.  \ref{fig:comp2}.

\begin{figure}
\centering
\includegraphics[width=6cm,angle=270]{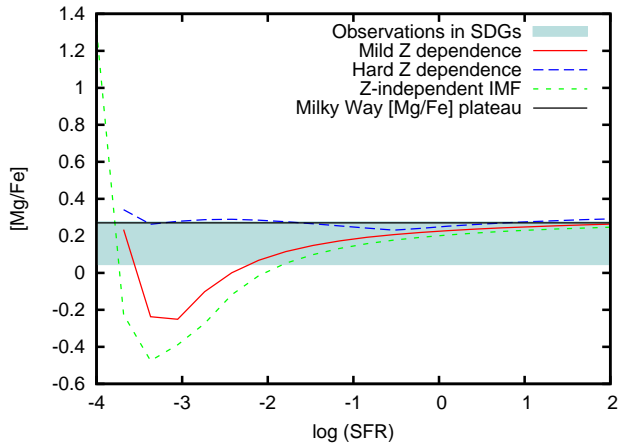}
\caption[]{\label{fig:mgfe_zdep} The variation of the [Mg/Fe] plateau
  as a function of the SFR for the mild (solid line) and hard
  (long-dashed line) dependencies of the IGIMF on the metallicity.
  Also the [Mg/Fe] predicted by the Z-independent IMF is shown
  (short-dashed line).  This is the same curve shown in Fig.
  \ref{fig:comparison}.  The average Milky Way [Mg/Fe] plateau as
  derived by Cayrel et al. (2004) is shown as a horizontal solid line
  (see Table \ref{table2}).  The range of variation of [Mg/Fe] in the
  considered DGSs is reported as a shaded area (see Table
  \ref{table4}). }
\end{figure}

We can repeat the calculations of Sec. \ref{sec:resgal} and evaluate
the dependence of the [$\alpha$/Fe] plateaus on the galaxy mass
assuming the mild and the hard IGIMF dependencies on [Fe/H].  We
consider only the reference values $\beta=2.00$ and ${\hat y}_{Fe}=0$.
Fig.  \ref{fig:mgfe_zdep} shows the [Mg/Fe] plateau vs. SFR plot for
the mild (solid line) and the hard (long-dashed line) models.  The
[Mg/Fe] calculated with the metallicity-independent IMF employed in
Sect. \ref{sec:resgal} is also shown for comparison (short-dashed
line).  The mild model shows a trend similar to the one of the
reference model, although the minimum [Mg/Fe] is up to $\simeq$ 0.2
dex above the one shown by the reference (metallicity-independent)
model.  This difference reduces considerably for larger values of SFR.
On the other hand, the hard model shows quite a different pattern: the
variations of [Mg/Fe] as a function of SFR are of modest magnitude and
[Mg/Fe] remains almost constant at $\sim$ 0.3 dex.
This means that low mass galaxies could indeed have [$\alpha$/Fe]
plateau values similar to the MW ones.  In Fig.  \ref{fig:mgfe_zdep}
we also plotted (shaded area) the data about the [Mg/Fe] ratios of the
most metal-poor stars in Draco, Carina, Sculptor and Sextans, putting
together all the calculated averages (with sigma deviations) for the
four DGSs (see Table \ref{table4}).  The curve relative to the hard
model is always close to the upper edge of this area, indicating that
the hard model can be compatible with the Milky Way [$\alpha$/Fe]
plateau and predicts [Mg/Fe] ratios that are only slightly larger than
those of the most metal-poor stars in DGSs.  On the other hand, the
mild model is always closer to the shaded area than the Z-independent
IGIMF model.  It appears to be compatible with observations provided
that the SFR is larger than $\sim$ 6 $\cdot$ 10$^{-3}$ M$_\odot$
y$^{-1}$ or smaller than $\sim$ 4 $\cdot$ 10$^{-4}$ M$_\odot$
y$^{-1}$.

It is interesting to understand why the [$\alpha$/Fe] plateau remains
almost constant as SFR increases in the hard model.  On the one hand,
galaxies with low SFR have, according to the fundamental metallicity
relation, low metallicities, hence flat IGIMFs (see Fig.
\ref{fig:comp2}).  On the other hand, galaxies characterised by low
SFRs attain a quite low value of $M_{\rm ecl, max}$ (see Eq.
\ref{eq:igimf}) and, consequently, a more limited amount of massive
stars.  These two competing effects (nearly) cancel out in this
formulation.

\subsection{On the comparison between our model predictions and the
  observations}
\label{sec:compwithobs}

As we have already anticipated in Sect. \ref{sec:satellites} and seen
in Sect.  \ref{sec:resgal}, in spite of the enormous wealth of data on
spectroscopy of metal-poor stars in dwarf galaxies and on SF histories
of DGSs, detailed comparisons between our models and the observations
are still difficult.  Our results should be seen thus as {\it
  predictions} and detailed comparison with observations will be
possible only when more data will be available and literature results
will be more homogeneous.  However, we can already comment on the
results of our study on the light of the available
observations/knowledges of the DGSs.  We recall briefly from Sect.
\ref{sec:sfh_obs} that typical early SFRs in DGSs are deduced to be of
the order of 10$^{-4}$ M$_\odot$ y$^{-1}$ but, for some galaxies, they
could have been one or two orders of magnitude larger.  Some
theoretical studies even suggest them to be of the order of 10$^{-1}$
M$_\odot$ y$^{-1}$ or more (Salvadori et al. 2008).  As we have seen
in Sect. \ref{sec:resgal}, at this level of SF the predicted
[$\alpha$/Fe] plateaus are very close to the ones observed in the MW.
Such a SFR would make our computation of [Mg/Fe] fully compatible with
the observed plateaus in Sextans, Carina and Draco (see again Fig.
\ref{fig:comparison}).  On the other hand, large differences are
predicted if the SFR is significantly below 10$^{-2}$ M$_\odot$
yr$^{-1}$.

We have reported in Table \ref{table4} the ranges of acceptable SFRs
for each galaxy, according to the comparison between (Z-independent)
IGIMF model results and observations shown in Fig.
\ref{fig:comparison}.  We can see that Carina, Sextans and Sculptor
can have had early SFRs as low as 0.1 M$_\odot$ y$^{-1}$ or less,
whereas Draco is the only considered DGS requiring larger SFRs (at
least 0.3 M$_\odot$ yr$^{-1}$).  Also Fornax, with a [Mg/Fe] plateau of
$\sim$ 0.1 dex (see Sect. \ref{sec:afe_obs}), is compatible with an
early SFR of $\sim$ 0.04 M$_\odot$ yr$^{-1}$.  These SFRs are in
general larger than the ones found in the literature for DGSs and
summarised above.  We remark again that, according to Fig.
\ref{fig:comparison}, the [Mg/Fe] ratios of the most metal-poor stars
in DGSs could be compatible also with a very low SFR (less than
10$^{-3}$ M$_\odot$ y$^{-1}$).  However, we believe that this solution
is unlikely because, in this interval of SFRs, the average [Mg/Fe]
ratios vary strongly.  The average [Mg/Fe] ratios in the four
considered DGSs are however quite similar (Table \ref{table4}) and
this fact seems to be hard to reconcile with the derived strong
dependence of [Mg/Fe] with SFR.  The SFRs suggested by the models
(Table \ref{table4}) and the ones available from the literature seem
thus to be different.  The reasons for this mismatch could be:
\begin{itemize}
\item As we have already argued, the determinations of the SFR in DGSs
  based on synthetic CMDs do not correctly take into account the stars
  lost during the DGS orbit around the Milky Way and, thus,
  under-estimate the intrinsic early SFR.
\item The IMFs in galaxies are indeed metallicity-dependent.  As we
  have shown, both mild and hard models predict [Mg/Fe] plateaus that
  are closer to the observations than what predicted by the
  Z-independent IGIMF model (see Fig.  \ref{fig:mgfe_zdep}).
\item At variance with our assumptions, the IMF in galaxies is indeed
  universal and, therefore, the [$\alpha$/Fe] plateaus of galaxies
  must always be the same (see also Kirby et al. 2011).
\end{itemize}
Based on the presently known observational facts, none of these three
possibilities can be definitely ruled out.  As described in Sect.
\ref{sec:resgal}, available observations for Draco, Carina, Sextans
and Sculptor suggest that the most metal-poor stars in these DGSs have
apparently slightly lower average [Mg/Fe] ratios than the most
metal-poor Milky Way stars.  This is an indication that the IMF can
indeed be non-universal.  However, once again one should not forget
that other high-quality data in other DGSs (Frebel et al.  2010,
Norris et al. 2010, Simon et al. 2010, Vargas et al. 2013) indicate
the presence of very metal-poor stars with [$\alpha$/Fe] ratios equal
or even larger that the Cayrel plateau.  Clearly, these data also
demand a theoretical explanation.  We can at the present not even
exclude the possibility that [$\alpha$/Fe] ratios of the most
metal-poor stars in some (or most of) DGSs of the Milky Way are larger
than the MW plateau value.  In this case, our results could be
compatible with the observations only if we consider the ``hard''
Z-dependent IMF model (see again Fig. 8).

Again, our results should be seen as predictions and, once data
concerning SFRs and [$\alpha$/Fe] plateaus in dwarf galaxies will be
more reliable, we will be able to establish more precisely if and to
what extent metallicity affects the slopes of the IMF in galaxies.

Another hypothesis could be considered here.  Some authors (see Kroupa
2012, Kroupa et al. 2010 and references therein) find that DGSs cannot
be of cosmological origin.  At least a fraction of them can be
originated by the tidal disruption of a much larger galaxy (see also
Lynden-Bell 1976).  A key probe of that is the alignment of most of
the DGSs along a disk (the {\it disk of satellites}), almost
perpendicular to the MW disk.  Another spectacular example of disk of
satellites has been recently discovered around Andromeda (Ibata et al.
2013).  If this is the case, of course the abundance patterns of the
metal-poor stars in these galaxies reflect the early conditions of
this large, disrupted galaxy.  The [$\alpha$/Fe] plateaus in the
fragments that emerge after the disruption of the parent galaxy should
be determined by the SFR in the early evolution of this primordial
galaxy.  They should therefore be quite similar to the MW plateau,
without big differences from galaxy to galaxy.  Of course, better
measurements of the [$\alpha$/Fe] ratios in DGSs and larger samples
can shed light on this scenario, too.

\section{Conclusions}
\label{sec:conc}

In this work we have studied the dependence of the [$\alpha$/Fe]
ratios in the most metal-poor stars of a galaxy (the [$\alpha$/Fe]
plateaus) on the galaxy's initial mass distribution of stars.  The
working hypothesis we have assumed in this study is that the most
metal-poor stars in a galaxy are made out of gas that has been solely
polluted by Type II SNe.  Ejecta from intermediate-mass stars and from
Type Ia SNe are released on longer timescales and cannot contribute to
the chemical enrichment at this early stage.  In this way we can
establish a connection between the high mass slope of the IMF and the
[$\alpha$/Fe] plateau.  If the high mass IMF slope is universal, than
the [$\alpha$/Fe] plateau cannot change from galaxy to galaxy.  If
instead different galaxies are characterised by different IMF slopes,
then a correlation between [$\alpha$/Fe] plateau and IMF could be
theoretically established.

In particular, the IGIMF theory (Kroupa \& Weidner 2003) predicts a
steep IMF in dwarf galaxies, characterised by mild SFRs.  One could
thus test the IGIMF theory by calculating the [$\alpha$/Fe] plateaus
as a function of the average SFR in a galaxy and compare them with
available observations in the best observed sample of dwarf galaxies,
namely the satellites of the Milky Way.  However, as summarised in
Sect.  \ref{sec:afe_obs}, neither the early SF histories nor the
[$\alpha$/Fe] plateau values can be firmly established for most of the
MW DGSs.  Our results should be seen as predictions of the IGIMF
models, to be compared with observations only once the determinations
of these quantities will be more reliable.

Moreover, recent works (Kroupa et al. 2013; M12) suggest that the IMF
in star clusters (hence also the integrated galactic IMF) could depend
on the star cluster's metallicity.  We have derived from M12 a mild
and a hard dependence of the IGIMF on the average galaxy metallicity
and we have calculated the [$\alpha$/Fe] plateaus accordingly.

Our main results can be summarised as follows:
\begin{itemize}
\item Dwarf galaxies characterised by low SFRs, i.e. SFRs smaller than
  10$^{-2}$ M$_\odot$ yr$^{-1}$, can show very low [O/Fe] or [Mg/Fe]
  plateaus, up to 0.7 dex lower than the MW [O,Mg/Fe] plateau.  This
  is at variance with observations, indicating [Mg/Fe] plateaus in
  dwarf galaxies only moderately lower than the Milky Way [Mg/Fe]
  value.  On the other hand, at SFRs smaller than $\sim$ 10$^{-3.5}$
  M$_\odot$ yr$^{-1}$, the [O,Mg/Fe] plateau can be much larger and
  consistent with observations.
\item If the SFR of the dwarf galaxy is larger than 10$^{-2}$
  M$_\odot$ yr$^{-1}$, then the variations of the [$\alpha$/Fe]
  plateaus with SFR are of the order of 0.2 dex at most.
\item The [$\alpha$/Fe] plateaus depend on the (uncertain) stellar
  yields between 8 and 12 M$_\odot$.
\item The slope $\beta$ of the mass distribution function of star
  clusters has a significant impact, too, on the chemical abundance
  pattern.
\item Our study of the observed [Mg/Fe] plateau in local dwarf
  galaxies suggests that they should be characterised by early SFRs
  either of the order of 10$^{-4}$ M$_\odot$ yr$^{-1}$ or larger than
  10$^{-2}$ M$_\odot$ yr$^{-1}$.  However, in the former hypothesis,
  given the steep [Mg/Fe]-SFR relation predicted within the IGIMF, it
  seems difficult to conceive similar [Mg/Fe] ratios in different
  galaxies as observed today.  It is also worth stressing that current
  determinations of the early SFRs in dwarf galaxies tend to be lower
  than 10$^{-2}$ M$_\odot$ yr$^{-1}$.  We have argued though that
  these SFR estimate are necessarily lower limits because they can not
  take into account stars stripped off the main body of the dwarf
  galaxy by tidal interactions.
\item For the first time, we have considered the effects of a
  metallicity dependent IMF on the chemical abundance ratios.  In case
  of a weak dependence of the IMF on [Fe/H], our results do not change
  substantially from the ones of the Z-independent IMF.  The minimum
  of [Mg/Fe] as a function of the SFR lies $\simeq$ 0.2 dex above the
  corresponding minimum for the standard, non metallicity-dependent
  IGIMF model.  For this model, there is agreement between model
  predictions and observations provided that the early star formation
  in the dwarf galaxies is larger than $\sim$ 6 $\cdot$ 10$^{-3}$
  M$_\odot$ y$^{-1}$ or smaller than $\sim$ 4 $\cdot$ 10$^{-4}$
  M$_\odot$ y$^{-1}$.
\item If instead the IMF strongly depends on the metallicity (hard
  model), galaxy models characterised by low SFRs will have flatter
  IGIMFs than galaxy models with high SFRs.  Unexpectedly, almost no
  dependence of the [$\alpha$/Fe] plateau on the SFR is predicted in
  this case.  The predicted [Mg/Fe] ratio is very close to the one
  observed in the most metal-poor stars of the Milky Way and only
  slightly larger than that observed in Local Group dwarf galaxies.
\end{itemize}

The enormous progress made recently in the determination of the star
formation histories and of the [$\alpha$/Fe] ratios of the most
metal-poor stars in local dwarf galaxies offers a valuable benchmark
for chemical evolution studies.  The next few years will see an
increase of the samples of stars with observable abundances in
low-mass, low metallicity galaxies.  With improved statistics and more
homogeneous data, our predictions for the variation of the
[$\alpha$/Fe] plateaus as a function of the SFR, will represent
valuable tools to constrain the most basic chemical evolution
parameters of dwarf galaxies, such as the IMF and the star formation
history.

To conclude, given the uncertainties related to the early star
formation rates in dwarf galaxies, we believe there are two ways to
better assess the reliability or our results: $(i)$: A better and more
precise determination of the [$\alpha$/Fe] plateaus in satellite dwarf
galaxies, to check whether these plateaus, at least for some galaxies,
could really differ from the MW halo values.  $(ii)$ A more direct
determination of the IMF in low-metallicity dwarf galaxies (see e.g.
Zhang et al. 2007) which could be directly used as an input for our
models for calculating abundance ratios of metal-poor stars in DGSs.

\section*{Acknowledgements}
We thank the anonymous referee for important suggestions and
constructive criticisms which improved the clarity and quality of the
paper. BKG, SR, and RC acknowledge the support of the UK's Science \&
Technology Facilities Council (ST/F002432/1 \& ST/H00260X/1). BKG
acknowledges the generous visitor support provided by Saint Mary's
University and Monash University.  SR acknowledges the kind
hospitality of the Jeremiah Horrocks Institute, University of Central
Lancashire, where this work has been first conceived.

\appendix
\section{Widening the parameter space}
\label{sec:app}
In this section, we extend the analysis made in Sect. \ref{sec:resmw}
and \ref{sec:resgal} and explore further parameters that might change
our results concerning [$\alpha$/Fe] plateaus in galaxies in the
framework of the IGIMF theory.  A summary of the results included here
is also presented in Table \ref{table3}.

As reported in Table \ref{table2}, the reference threshold mass for
SNII explosion used in Eq. \ref{eq:plateau} is 8 M$_\odot$.  If we
increase it to 10 M$_\odot$, the [$\alpha$/Fe] variations as a
function of ${\hat y}_{Fe}$ become very weak ($\sim$ 0.04 dex).  This
was expected because, although the fraction of massive stars with
masses between 10 and 12 M$_\odot$ is not negligible ($\sim$ 16 \% for
$\beta=2.00$), the heavy element production in this mass range is
quite limited.  For instance, only $\sim$ 1.8\% of oxygen is produced
by stars in the interval [10,12] M$_\odot$ for $\beta=2$ and ${\hat
  y}_{Fe}=0.5$.  Conversely, we can observe that, modifying the
standard value of $m_{thr}$ from 8 to 10 M$_\odot$, we also obtain
very small changes (of the order of 0.04 dex) in the [$\alpha$/Fe]
ratios.

The value of the average MW SFR in the first Gyr of 0.5 M$_\odot$
yr$^{-1}$, adopted as a standard value, comes from the C10
simulations.  This value is however both model- and
assumption-dependent.  Significantly different (usually larger) values
can be found in the literature.  For instance Gibson et al. (2008)
show different simulations of the evolution of MW-like galaxies and
the average SFR in the early phases is much larger than 0.5 M$_\odot$
yr$^{-1}$, reaching SFRs up to $\simeq$ 20--30 M$_\odot$ yr$^{-1}$.
We have repeated the calculations of the [$\alpha$/Fe] ratios using a
SFR of 100 M$_\odot$ yr$^{-1}$, the largest SFR considered by R09.
The results differ very little from the ones presented in Fig.
\ref{fig:fit2}: the [O/Fe] ratios are at most 0.015 dex larger than
the reference values.  This confirms the general conclusion of R09
that only for SFRs below 0.1--0.01 M$_\odot$ yr$^{-1}$ are the
deviations of the IGIMF from a standard (not SFR-dependent) IMF
significant.

Yields above 40 M$_\odot$ turn out to affect negligibly our results.
Even if we increase the iron yields above 40 M$_\odot$ by a factor of
10$^4$, the [$\alpha$/Fe] ratios change by less than 0.01 dex (see
Table \ref{table3}).  We have also varied the parameters $A$ and $B$
characterising the dependence of the maximum embedded cluster mass
$M_{\rm ecl, max}$ and the SFR in Eq.  \ref{eq:igimf}.  A variation of
$A$ of the order of 0.5 and of $B$ of the order of 0.1 change very
little our results.  In fact, [O/Fe] changes by $\sim$ 0.05 dex at
most.  We remind however that the correlation between $M_{\rm ecl,
  max}$ and $\psi$ is based on observations, therefore it should not
be subject to large variations.  

We have varied also the assumed lower mass for an embedded cluster
$M_{ecl, min}=$ 5 M$_\odot$ and the theoretical upper stellar mass
within an embedded cluster $m_{max *}=$ 150 M$_\odot$ (see Table
\ref{table1}) that are used in the IGIMF formula (Eq.
\ref{eq:plateau}).  Since the upper end of the IGIMF is affected by
the most massive star cluster, not by the smallest ones, assuming
$M_{ecl, min}$ larger than 5 M$_\odot$ produces negligible effects.
For instance, if we take $M_{ecl, min}=20$ M$_\odot$ we obtain results
differing by less than $\sim$ 0.01 dex compared to the reference
models.  Since the work of Crowther et al. (2010) indicates the
existence of stars with masses larger than 150 M$_\odot$, it is worth
studying what changes in our results if we assume $m_{max *}=$ 300
M$_\odot$.  At low SFRs the results with $m_{max *}=$ 300 M$_\odot$
are practically indistinguishable from the ones we have presented so
far.  This is understandable because in this case the upper mass
predicted by the IGIMF theory is much below 150 M$_\odot$.  At large
SFRs, some differences start emerging, but they never exceed $\sim$
0.03--0.04 dex.  Again, these results are summarised in Table
\ref{table3}.

\end{document}